\documentclass[aps,pra,reprint,superscriptaddress]{revtex4-1}
\pdfoutput=1

\usepackage{epsfig}
\usepackage{graphicx}
\usepackage{dcolumn}
\usepackage{amsmath}
\usepackage[english]{babel}

\begin{document}
\title{How van der Waals interactions determine the unique properties of water}

\author{Tobias Morawietz}
\email[]{tobias.morawietz@theochem.ruhr-uni-bochum.de}
\email[Present address: ]{Faculty of Physics, University of Vienna, A-1090 Vienna, Austria.}
\affiliation{Lehrstuhl f\"ur Theoretische Chemie, Ruhr-Universit\"at Bochum, D-44780 Bochum, Germany.}
\author{Andreas Singraber}
\author{Christoph Dellago}
\affiliation{Faculty of Physics, University of Vienna, A-1090 Vienna, Austria.}
\author{J\"org Behler}
\email[]{joerg.behler@theochem.ruhr-uni-bochum.de}
\affiliation{Lehrstuhl f\"ur Theoretische Chemie, Ruhr-Universit\"at Bochum, D-44780 Bochum, Germany.}

\begin{abstract}
While the interactions between water molecules are dominated by strongly directional hydrogen bonds (HBs), it was recently proposed that relatively weak, isotropic van der Waals (vdW) forces are essential for understanding the properties of liquid water and ice.
This insight was derived from \textit{ab initio} computer simulations, which provide an unbiased description of water at the atomic level and yield information on the underlying molecular forces.
However, the high computational cost of such simulations prevents the systematic investigation of the influence of vdW forces on the thermodynamic anomalies of water.
Here we develop efficient \textit{ab initio}-quality neural network potentials and use them to demonstrate that vdW interactions are crucial for the formation of water's density maximum and its negative volume of melting.
Both phenomena can be explained by the flexibility of the HB network, which is the result of a delicate balance of weak vdW forces, causing e.g. a pronounced expansion of the second solvation shell upon cooling that induces the density maximum.
\end{abstract}

\maketitle

Water is an exceptional liquid, exhibiting several anomalies of which the density maximum at 4\,$^{\circ}$C is the most prominent one~\cite{Ludwig2001}. Together with the negative volume of melting, it is responsible for the fact that water freezes from the top down and ice floats on water. The unusual behavior of water can be directly related to its ability to form hydrogen bonds (HBs) which are of strongly directional nature and determine the microscopic structure of water~\cite{Stillinger1980,Mishima1998}. To investigate the anomalies of water at the molecular level atomistic computer simulations have become an essential tool complementary to experimental studies. Such simulations are applicable even at conditions not accessible in experiment~\cite{Mishima1998,Palmer2014} and important contributions have been made by simulations employing simple empirical water models~\cite{Mishima1998,Palmer2014,Jorgensen1983,Poole1992,Paschek2005,Poole2005}.

Simulations based on \textit{ab initio} molecular dynamics (AIMD)~\cite{Car1985,Marx2009,Hassanali2014} allow to determine the properties of water with high predictive power and enable a detailed analysis of their underlying microscopic mechanisms. In contrast to empirical water models~\cite{Jorgensen1983}, which depend on experimental data resulting in a limited transferability, in AIMD the atomic forces that govern the molecular dynamics are obtained directly from quantum mechanics. While this approach is in principle exact (in combination with methods that account for the quantum nature of the nuclei~\cite{Morrone2008,Ceriotti2013}), \textit{ab initio} simulations of condensed matter systems are feasible only if approximate but efficient methods such as density-functional theory (DFT) are employed. Even then, however, simulations are restricted to short times and small systems.
AIMD simulations have been employed to a limited extent to investigate the phase behavior of water, for instance by estimating melting temperatures~\cite{Yoo2009,Yoo2011} and vapor-liquid coexistence curves~\cite{McGrath2006,McGrath2006a}. However, many fundamental thermodynamic properties of water have not been evaluated to date.
In order to circumvent the limitations of on-the-fly AIMD, various efficient water potentials employing data from \textit{ab initio} calculations have been developed. For instance, existing water models have been re\-pa\-ra\-me\-tri\-zed, based solely on forces from AIMD simulations~\cite{Spura2015} or using a combination of experimental and theoretical data~\cite{Wang2014}. 
Other potentials employ truncated many-body expansions of the water interaction energy, with parameters that are fitted to \textit{ab initio} results for small water clusters ~\cite{Wang2010,Bartok2013,Medders2014}.
Recently, is was shown that a minimal water model with a coarse grained electronic structure described by quantum Drude oscillators~\cite{Jones2013} (QDOs) is able to predict many thermodynamic properties of water~\cite{Sokhan2015}.

Here, we present a series of analytic potentials which accurately represent the \textit{ab initio} potential-energy surface of water and overcome the computational bottleneck of AIMD simulations, enabling to assess the performance of different density-functionals. The form of the potentials is not constructed employing simplified physically motivated models, but instead consists of a set of highly flexible functions in form of artificial neural networks~\cite{Behler2007,Behler2014} trained to a broad range of condensed phase configurations. Using this powerful approach we carry out converged large-scale molecular dynamics simulations of water and clarify the significance of vdW interactions for the thermodynamic anomalies of water.

\section{Results and Discussion}

\subsection{Neural Network Potentials}

We developed four neural networks potentials (NNPs) representing the RPBE~\cite{Hammer1999} and BLYP~\cite{Becke1988,Lee1988} density-functionals with and without vdW corrections employing the D3 method~\cite{Grimme2010}. While NNPs can in principle be trained to any reference method, the majority of AIMD simulations for water reported to date have employed gradient corrected (GGA) DFT, and RPBE and BLYP are two well-established density-functionals within this class.

\begin{figure}[htbp]
\centering
  \includegraphics{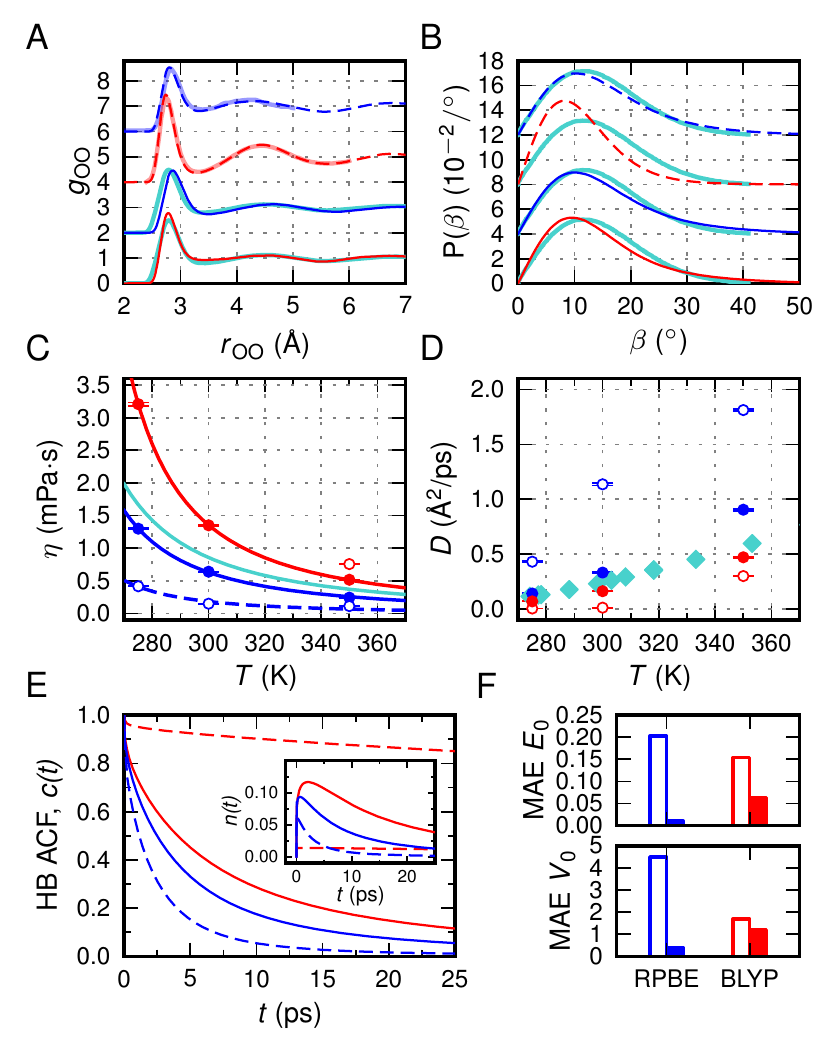}\\  
\caption{
\label{fig:fig01}
\textbf{Static and dynamic properties of liquid and crystalline water.}
(A-B) Structure of water at $T$\,=\,$300$\,K obtained from NNP simulations based on RPBE (dark blue lines) and BLYP (dark red lines) density-functionals with (solid lines) and without (dashed lines) vdW corrections, compared to data from AIMD~\cite{Fernandez-Serra2005,Morrone2008} (light blue and red lines) and experimental results~\cite{Soper2013,Modig2003} (cyan lines). (A) Oxygen-oxygen radial distribution functions \textit{g}$_{\text{OO}}$ and (B), probability density function P($\beta$) of the HB angle $\protect\beta=\protect\angle\protect\text{H}_\text{D}\protect\!\!-\protect\!\protect\text{O}_\text{D}\protect\!\protect\cdots\protect\!\protect\text{O}_\text{A}$. The curves are shifted for better visibility. (C-D) Viscosity $\eta$ and diffusion coefficient \textit{D} as function of temperature obtained from NNP simulations compared to experimental values~\cite{Yoshida2005,Bird2002} (cyan line / cyan diamonds). Diffusion coefficients are corrected for finite size effects (cf. Appendix, Fig.~\ref{fig:diffusion_viscosity}). Lines in (C) were obtained from a fit of the form: $\eta = (T - T_0)^{-b}$. (E) Hydrogen bond autocorrelation functions $c(t)$ and $n(t)$ (cf. Ref.~\cite{Luzar1996} and Appendix) obtained from NNP simulations at $T$\,=\,$300$\,K. (F) Mean absolute errors (MAEs) with respect to experiment of equilibrium lattice energies \textit{E}$_\text{0}$ (in eV/H$_2$O) and volumes \textit{V}$_\text{0}$ (in \AA{}$^3$/H$_2$O) of seven ice phases computed at $T$\,$=$\,0\,K with NNPs based on plain (empty bars) and vdW-corrected (filled bars) density-functionals. Energy vs. volume curves of all ice phases are reported in Appendix, Fig.~\ref{fig:ice_e_vs_v}.
}
\end{figure}

The parameter set of each NNP was obtained in an iterative procedure using energies and forces from periodic configurations of liquid and crystalline water under various conditions. Details concerning the functional form and the composition of the reference data set of the NNPs are given in the Appendix.
Root mean squared errors of energies and forces in the final NNP training sets are $\approx$\,2\,meV/H$_2$O and $\approx$\,70\,meV/\AA{}, respectively, and the error for configurations not included in the training set is of comparable order. Such errors are well below the intrinsic uncertainties of the DFT calculations related to the exchange-correlation functional and comparable to the much smaller error due to the finite basis set size. The NNPs closely reproduce the properties of liquid and crystalline water obtained from DFT calculations (cf Fig.~\ref{fig:fig01}a and Appendix, Figs.~\ref{fig:VDOS} and~\ref{fig:ice_e_vs_v}) and are thus well suited for assessing the quality of the underlying reference method. The inability of GGA density-functionals to describe vdW forces can be compensated by vdW correction schemes~\cite{Grimme2010}. Comparing simulations with and without correction then allows us to investigate the effect of vdW interactions on the properties of water.

\begin{figure}[htbp]
 \centering
  \includegraphics{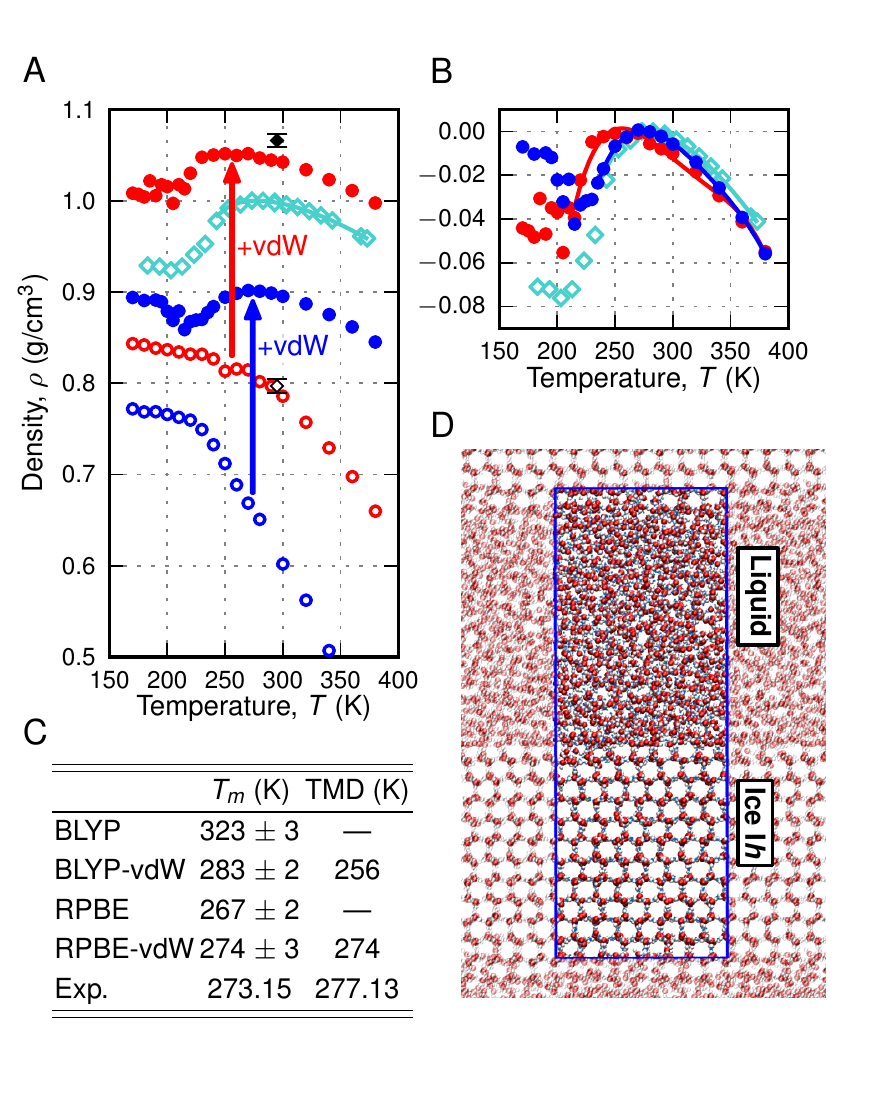}\\   
 \caption{\label{fig:fig02}\textbf{Density isobars and melting temperature.} (A) Density isobars at \textit{p}\,=\,1\,bar obtained from NNP simulations based on the BLYP (red empty circles), BLYP-vdW (red filled circles), RPBE (blue empty circles) and RPBE-vdW (blue filled circles) density-functionals. For comparison, experimental data for bulk (cyan line) and confined water~\cite{Mallamace2007} (cyan diamonds), and data from AIMD simulations with the vdW-corrected (black filled diamond) and uncorrected (black empty diamond) BLYP density-functional~\cite{DelBen2015} are also shown. (B) Densities relative to the density maximum. The density maxima for BLYP-vdW and RPBE-vdW were obtained from polynomial fits (red and blue lines) to the density isobars. (C) Melting temperature (\textit{T}$_\text{\textit{m}}$, errors were estimated by block averaging) and temperature of maximum density (TMD). Melting temperatures were corrected for deviations between the NNP and the DFT potential-energy surfaces (cf. Appendix). (D) Snapshot of coexisting liquid water and ice I\textit{h} taken from the interface pinning simulation used to determine the melting temperature (the simulation cell is drawn in blue).}
\end{figure}

The large impact of vdW forces on the properties of liquid and crystalline water as obtained from NNP simulations is illustrated in Fig.~\ref{fig:fig01}. VdW interactions soften the water structure, reducing deviations from the experimental curves, and significantly improve lattice energies and volumes of various ice polymorphs.
The influence on the dynamic properties depends on the respective density-functional. While the BLYP-based simulations exhibit a very low water mobility, the opposite behavior is observed for RPBE. Including vdW interactions improves both methods, resulting in a more realistic description of the dynamics of water. Similar conclusions have been drawn from AIMD simulations~\cite{Lin2009,Schmidt2009,Wang2011b,Ma2012,Forster-Tonigold2014,DiStasio2014,DelBen2015} and static DFT calculations~\cite{Santra2011,Brandenburg2015}. However, the specific influence of vdW interactions on the thermodynamic anomalies of water is still unclear. In order to understand the atomistic origin of these fundamental properties, we have carried out large-scale NNP simulations to determine the density isobar of water and the melting temperature of ice I\textit{h}.

\begin{figure}[htbp]
 \centering
  \includegraphics{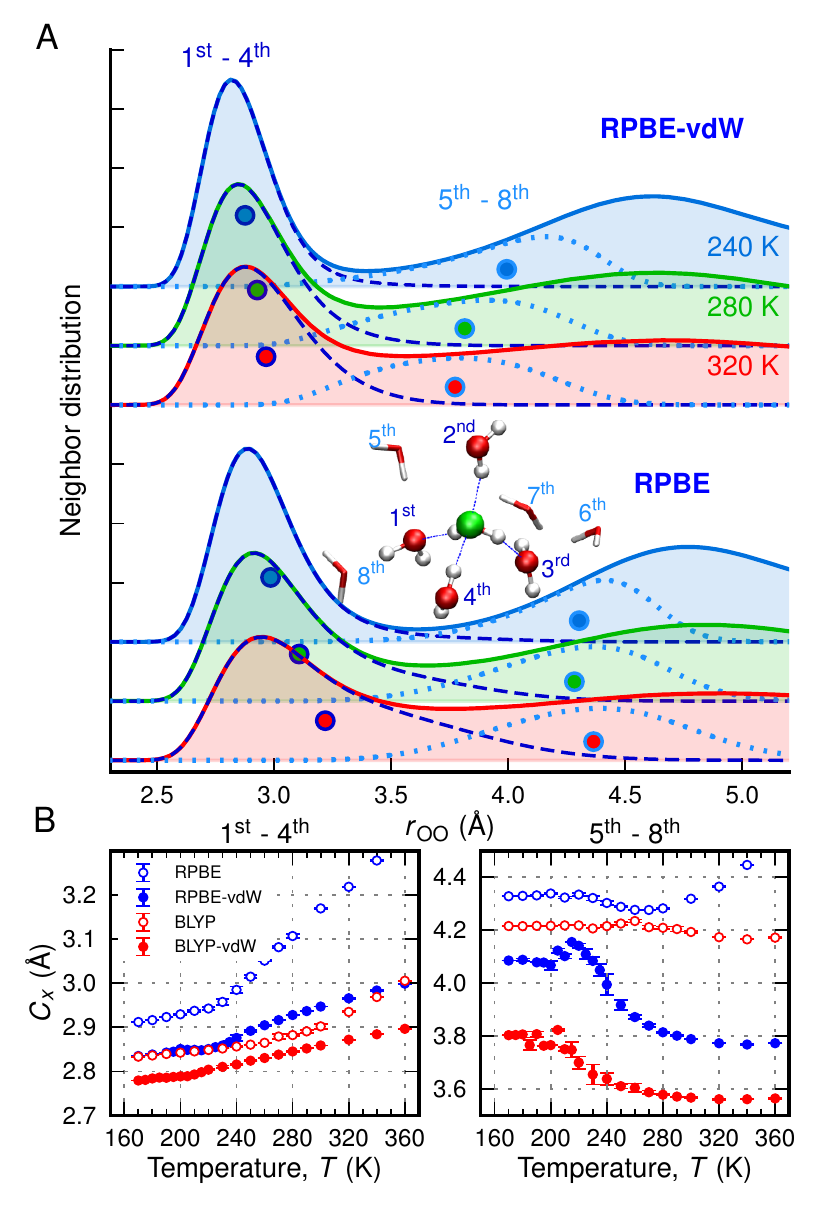}\\  
 \caption{\label{fig:fig03}\textbf{Temperature dependence of water neighbor distribution.}  (A) Oxygen-oxygen radial distribution functions (solid lines) and decompositions into contributions from molecules in the first shell (1$^{\text{st}}$ - 4$^{\text{th}}$ neighbors, dark blue dashed lines) and part of the second shell (5$^{\text{th}}$ - 8$^{\text{th}}$ neighbors, light blue dotted lines) together with the centroid of the neighbor distribution (dark blue and light blue circles) for three different temperatures from simulations with (RPBE-vdW, top panel) and without (RPBE, bottom panel) vdW corrections. 
(B) Position of the neighbor distribution centroid $C_x$ as function of temperature. In general, with decreasing temperature a contraction of the first shell is observed (left panel), while only in vdW corrected simulations a maximum of the total density is found due to the expansion of the second shell (right panel).
}
\end{figure}

\subsection{Density Isobars}
Fig.~\ref{fig:fig02}a compares density isobars down to the supercooled regime obtained from constant pressure simulations at \textit{p}\,=\,1\,bar to experimental data for bulk and confined water~\cite{Mallamace2007}. Simulations without vdW corrections show poor agreement with experiment: total densities are severely underestimated by 20\,--\,40\,\%, no density maximum is observed and the density increase upon cooling is too steep, resulting in strongly overestimated thermal expansivities at ambient conditions (cf. Appendix, Table~\ref{tab:density_max}). In contrast, the inclusion of vdW forces leads to qualitatively correct results: both vdW corrected potentials exhibit a density maximum, the shape of the isobars is in close agreement with experiment (cf. Fig.~\ref{fig:fig02}b) and total densities are shifted to larger values (in agreement with AIMD simulations performed at a single state point~\cite{Schmidt2009,DelBen2015,Gaiduk2015}), reducing deviations from experiment to 5\,--\,10\,\%. Moreover, a density minimum in the supercooled regime (at $\approx$200\,K and $\approx$215\,K for BLYP-vdW and RPBE-vdW, respectively) can be identified that is consistent with experimental measurements of water confined in silica pores~\cite{Mallamace2007} and simulations using empirical water models~\cite{Paschek2005,Poole2005}. 
Comparable conclusions regarding the role of vdW forces in determining the density profile of water have been obtained from simulations with modified empirical water models: Truncating Lennard-Jones interactions resulted in reduced densities (15\,\% lower compared to the full model) and the disappearance of the density maximum~\cite{Remsing2011}. Similar underestimated densities were found in simulation with the QDO water model, where the experimental value could be retained by increasing the strength of many-body dispersion~\cite{Jones2013}.

\subsection{Melting Temperatures}
Exploiting the efficiency of NNPs, we have, for the first time, accurately computed the melting point of ice from first principles. Melting temperatures of ice I\textit{h} obtained using the interface pinning method~\cite{Pedersen2013} are listed in Fig.~\ref{fig:fig02}c. While the melting point is overestimated by about 50\,K in BLYP-based simulations, all other potentials (based on BLYP-vdW, RPBE and RPBE-vdW) agree within 10\,K with experiment. However, only when vdW interactions are accounted for, liquid water is denser than ice at coexistence (cf. Appendix, Table~\ref{tab:density_liquid_solid}) and the anomalous melting behavior that causes water to freeze from the top down can be reproduced.
Estimates of the melting temperature of water obtained earlier from AIMD simulations employing the BLYP density-functional have been reported to be 360\,K with~\cite{Yoo2009} and 411\,K without~\cite{Yoo2011} vdW corrections. While the lowering of the melting point of about 50\,K with inclusion of vdW forces is consistent with our results, the previous reported melting temperatures are much higher (for plain BLYP these high melting temperatures have been inconsistent with the calculated boiling point, which was estimated to be about 350\,K~\cite{McGrath2006}). This discrepancy may be attributed to the limited system size (192 molecules) and simulation time (15\,ps) of the AIMD simulations and the fact that they have been carried out at the experimental density of 1\,g/cm$^3$ rather than at constant pressure.
While nuclear quantum effects (NQEs) not included here tend to weaken hydrogen bonds~\cite{Soper2008} and soften the structure of liquid water~\cite{Morrone2008} they are unlikely to qualitatively change our findings for the thermodynamic properties of water. Melting temperatures are only weakly affected due to a competition between intra- and intermolecular NQEs~\cite{Habershon2009,Markland2012,Romanelli2013}. Further, simulations with \textit{ab initio}-based potentials indicate that NQEs do not alter the location of the density maximum~\cite{Paesani2007} and only marginally reduce the absolute water density~\cite{Paesani2007,Medders2014}.

\subsection{Water Neighbor Distribution}
In order to identify the molecular origin of water's complex density isobar (exhibiting both a maximum and minimum) we have analyzed the structure of water by decomposing~\cite{Saitta2003} the oxygen-oxygen radial distribution function into contributions from first and second shell molecules (Fig.~\ref{fig:fig03}). At high temperatures, thermal fluctuations weaken HBs and lead to an increased distance to hydrogen bonded molecules in the first solvation shell. At the same time, second-shell molecules are able to penetrate the first shell (becoming interstitial molecules~\cite{DiStasio2014,Saitta2003,Jedlovszky2000}) and perturb the local tetrahedral water network. Upon cooling, the HB strength increases and the first shell approaches the central molecule (Fig.~\ref{fig:fig03}b, left panel) causing a density increase. However, this effect is compensated by a reduced number of interstitial molecules resulting in a shift of the second shell to larger distances, thus lowering the density (Fig.~\ref{fig:fig03}b, right panel) and inducing a density maximum.  Decreasing the temperature further, the expansion of the second shell finally saturates and the density increases again after passing through a minimum in the supercooled regime. Simulations without vdW corrections do not feature a pronounced second-shell shift, which explains the monotonic density increase with decreasing temperature in this case.

\begin{figure}[htbp]
 \centering
  \includegraphics{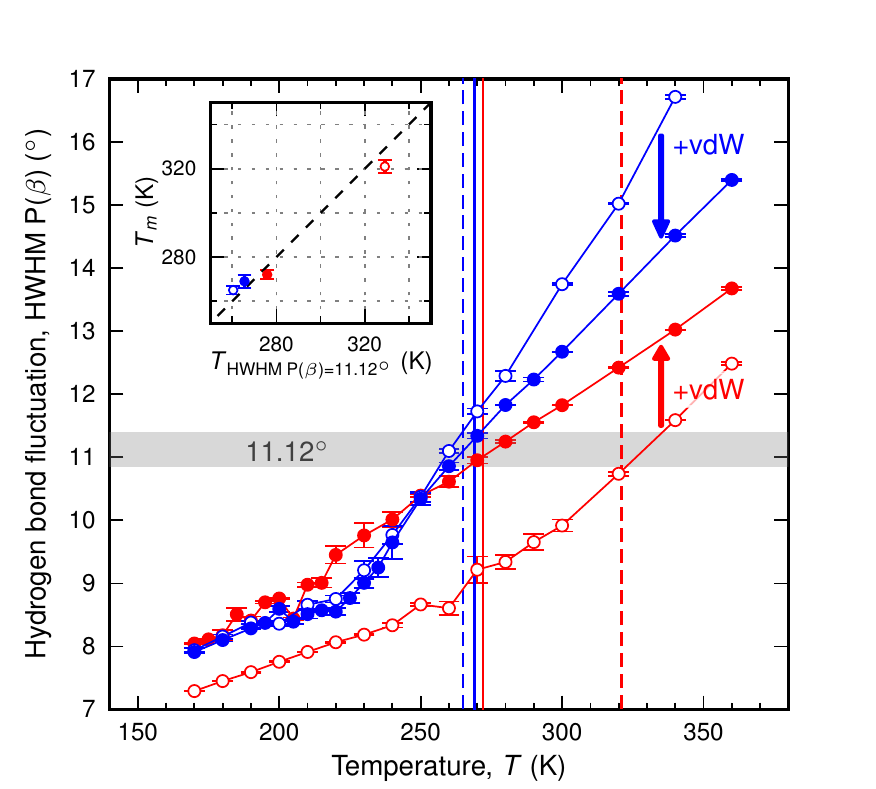}\\ 
 \caption{\label{fig:fig04}\textbf{Hydrogen bond fluctuation and melting temperature.} Fluctuation in the HB angle $\beta$, measured by computing the half width at half maximum (HWHM) of the distribution function P($\beta$), as function of temperature obtained from NNP simulations based on the BLYP (red empty circles), BLYP-vdW (red filled circles), RPBE (blue empty circles) and RPBE-vdW (blue filled circles) density-functionals. 
The vertical lines denote the melting temperature of the respective potentials (solid lines: vdW corrected potentials, dashed lines: uncorrected potentials). The inset shows the correlation between melting temperature \textit{T}$_\text{\textit{m}}$ and temperature of a HB fluctuation of 11.12$^\circ$. The error bars of the melting temperatures were obtained by block averaging.
}
\end{figure}

\subsection{Hydrogen Bond Analysis}
The absence of the density maximum in the simulations without vdW corrections can be understood by examining the strength of the HB network. In Fig.~\ref{fig:fig04} we analyze the fluctuations in the HB angle $\beta$, a measure of the HB strength, as function of temperature. 
The different magnitudes of HB fluctuations indicate that HBs in uncorrected simulations are either too strong (BLYP) or too weak (RPBE) to yield a density maximum. In BLYP-based simulations the tetrahedral water network is too rigid, so that even in the high temperature regime second-shell molecules are not able to penetrate the first solvation shell and a shift of the second shell at lower temperatures is prevented. HBs in RPBE-based simulations, on the other hand, are very weak, as manifested by large fluctuations in $\beta$ and large distances of first-shell molecules, so that the first and second solvation shells are both shifted to smaller distances upon cooling until at $\approx$260\,K HBs are strong enough to slightly reduce the number of interstitial molecules. This effect leads to a strong contraction of the system, resulting in high thermal expansivities (cf. Appendix, Table~\ref{tab:density_max}) with an almost 10-fold increase over the experimental value. These results are in line with the observations made for the dynamic properties of water shown in Fig.~\ref{fig:fig01}c-e.

Interestingly, we find that the HB fluctuations evaluated for the liquid phase can be used as a measure for the melting temperature of ice I\textit{h}, \textit{T}$_\text{\textit{m}}$, in analogy to the Lindemann melting rule~\cite{Lindemann1910}. For all NNPs employed here, ice I\textit{h} melts when the HB fluctuations exceed a critical value of $\approx$ 11$^\circ$ (cf. Fig.~\ref{fig:fig04}), which explains the different values for \textit{T}$_\text{\textit{m}}$ reported in Fig.~\ref{fig:fig02}c. In order to verify the correlation between HB strength and melting temperature, we performed simulations close to the melting temperature using a series of empirical water models. As shown in Appendix, Fig.~\ref{fig:hbond_melt}, the critical fluctuation value at which melting occurs depends slightly on the class of water model. For TIP4P~\cite{Jorgensen1983}-based models, a high degree of correlation is found at a critical value of $\approx$ 10$^\circ$.

\begin{figure}[htbp]
 \centering
  \includegraphics{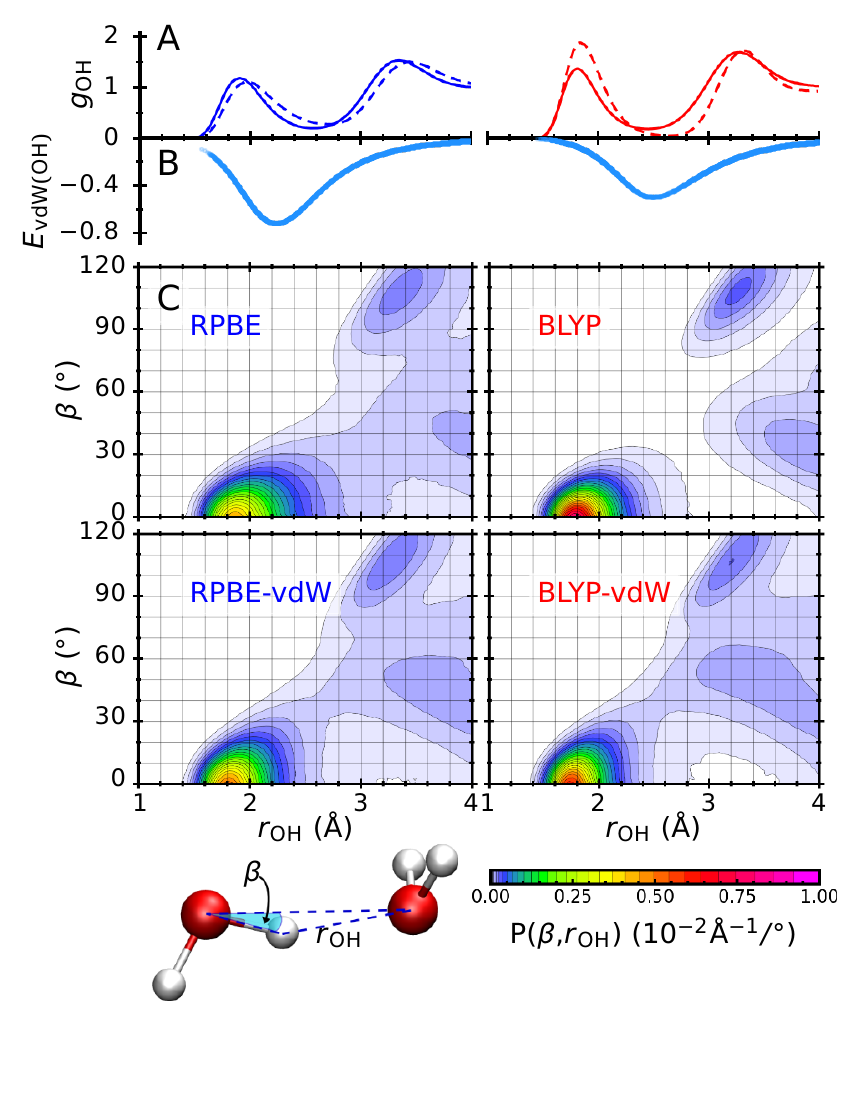}\\  
 \caption{\label{fig:fig05}\textbf{Effect of van der Waals interactions on hydrogen bond geometry.} (A) Intermolecular oxygen-hydrogen radial distribution functions \textit{g}$_\text{OH}$ from NNP simulations at 300\,K based on the RPBE (left) and BLYP (right) density-functionals with (solid lines) and without (dashed lines) vdW corrections. 
 (B) vdW pair interaction energy \textit{E}$_{\text{vdW(OH)}}$ (in $k_{\rm B}T$) 
 between oxygen and hydrogen atoms as a functions of the pair distance \textit{r}$_{\text{OH}}$ obtained from the NNP simulations. (C) Probability density function P($\beta$,\textit{r}$_{\text{OH}}$) of the HB angle $\beta$ and the oxygen-hydrogen distance \textit{r}$_{\text{OH}}$.
}
\end{figure}

The fact that vdW corrections either weaken (BLYP) or strengthen (RPBE) HBs is related to the vdW correction terms which are different for the two density-functionals (see Appendix, Table~\ref{tab:vdW_coeffs}). For RPBE, the vdW interactions between pairs of oxygen and hydrogen atoms (Fig.~\ref{fig:fig05}b, left) have a deep minimum at short OH distances increasing the HB strength by reducing the probability of configurations with extended intermolecular OH distances (cf. Fig.~\ref{fig:fig05}a, left). In contrast, the BLYP vdW pair interaction (Fig.~\ref{fig:fig05}b, right) is weaker and shifted to larger distances, inducing an increased population of extended HBs with reduced strength (Fig.~\ref{fig:fig05}a, right). Similar observations can be made for vdW interactions between pairs of oxygen atoms, shown in Appendix, Fig.~S10. Both effects are clearly visible in the probability density functions of HB angle and OH distance (Fig.~\ref{fig:fig05}c): HBs are very flexible in case of RPBE and very stiff in case of BLYP. Both vdW corrected density-functionals show similar distributions in between the uncorrected probability density functions.

\subsection{Dielectric properties}

We conclude by analyzing the influence of vdW interactions on the dielectric properties of liquid water. Calculating the dielectric constant from MD simulations requires simulation times of several nanoseconds. While rough estimates based on short AIMD trajectories have been made~\cite{Silvestrelli1999,Sharma2007}, converged values have not been obtained to date. We have computed converged dielectric constants at 300\,K by generating independent configurations in NNP simulations, which we then used to calculate molecular dipole moments from maximally localized Wannier functions~\cite{Marzari1997} (see Appendix). While the dielectric constant is overestimated in simulations without vdW interactions (159.8 and 108.9 for BLYP and RPBE, respectively) compared to the experimental value of 78.4, the values improve if vdW forces are accounted for (106.2 and 95.2 for BLYP-vdW and RPBE-vdW, respectively) which is associated with a shift of the molecular dipole moment distributions to lower values (see Appendix, Fig.~\ref{fig:dielectrics}a) in agreement with results from AIMD simulations~\cite{DiStasio2014}.

\subsection{Conclusions}
In summary, we have shown that \textit{ab initio}-based NNP simulations are able to correctly predict water's density maximum and melting temperature provided that vdW forces are taken into account, highlighting the important role of these interactions in determining the properties of aqueous systems. Despite their rather weak and anisotropic nature, vdW forces crucially modulate the HB network and ensure the right level of flexibility that causes the anomalous behavior of water. These results demonstrate the predictive power of AIMD simulations compared to empirical water models (whose density maximum is usually underestimated if experimental density information is not explicitly included in the parametrization process~\cite{Vega2005}). They further indicate that liquid water and ice can be accurately described already on the GGA level of theory if vdW corrections are considered without the need to include exact exchange in form of computationally demanding hybrid density-functionals. NNP simulations of water are thus a reliable and predictive tool which could also help investigating other important properties such as the existence of a liquid-liquid transition in water at supercooled conditions~\cite{Poole1992,Palmer2014}. Unlike most water models, NNPs are capable of describing the making and breaking of chemical bonds, opening the possibility to study proton transfer and autoionization in the condensed phase.

\section{Methods}

MD simulations were performed with an extended version of the LAMMPS program~\cite{Plimpton1995} using four \textit{ab initio}-based NNPs for water representing RPBE~\cite{Hammer1999} and BLYP~\cite{Becke1988,Lee1988} density-functionals with and without vdW corrections employing the D3 method~\cite{Grimme2010}. Parameter sets for all NNPs are publicly available~\cite{figshare2015}. 

Distribution functions compared to previous AIMD results (Fig.~\ref{fig:fig01}a, top two curves) were computed in the canonical (\textit{NVT}) ensemble at a density of 1\,g/cm$^3$, while for a comparison with experiment (Fig.~\ref{fig:fig01}a, bottom two curves and Fig.~\ref{fig:fig01}b) simulations were run in the isothermal-isobaric (\textit{NpT}) ensemble (see below). Dynamic properties (Fig.~\ref{fig:fig01}c-e) were obtained in the following way: for each NNP and temperature the equilibrium volume was determined by \textit{NpT} simulations as described below. Then, simulations in the \textit{NVT} ensemble were carried out for 1\,ns using 512 water molecules and a time step of 0.5\,fs. 32 statistically independent sets of coordinates and velocities were extracted from each \textit{NVT} trajectory and used as starting points for simulations in the microcanonical (\textit{NVE}) ensemble with a simulation time of 200\,ps per trajectory. After discarding the first 50\,ps for the purpose of equilibration, viscosities and diffusion coefficients corrected for finite size effects were determined (see Appendix for details). Hydrogen bond kinetics were analyzed in terms of the Luzar-Chandler model~\cite{Luzar1996} and computed with the GROMACS package~\cite{VanderSpoel2005,VanderSpoel2006}. The VDOS spectrum shown in Appendix, Fig.~\ref{fig:VDOS} was computed from velocity autocorrelation functions obtained from 16 independent \textit{NVE} simulations with a length of 20 ps using initial configurations from an \textit{NVT} trajectory at 300 K.

Density isobars at 1\,bar covering a temperature range from 380\,K to 170\,K in steps of 5\,-\,20\,K were obtained from molecular dynamics simulations of 128 water molecules in the \textit{NpT} ensemble with a time step of 0.5\,fs and employing the equations of motion of Shinoda et al.~\cite{Shinoda2004}. As shown in Appendix, Fig.~\ref{fig:system_size_density_isobar}, simulation cells containing 128 molecules are sufficient to obtain converged density isobars. The simulation length at each single temperature was 2\,-\,20\,ns (depending on temperature and convergence behavior), resulting in a total simulation time of more than 700\,ns. Configurations from equilibrated simulations in the \textit{NVT} ensemble were used as starting points for the \textit{NpT} simulations at 380\,K. Subsequently, simulations at lower temperatures were performed step by step by using the final configuration of the preceding simulation as initial configuration. The first halves of the trajectories served for equilibration and were not used for analyses. Density maxima and thermal expansivities at \textit{T}\,=\,25\,$^{\circ}$C were obtained from polynomial fits (4$^{\mathrm{th}}$-order for BLYP-vdW and RPBE-vdW, 3$^{\mathrm{rd}}$-order for BLYP and RPBE) to the density isobars.

Melting temperatures of ice I\textit{h} were computed employing the interface pinning method~\cite{Pedersen2013,Pedersen2013a} (see Appendix for details). The densities of the liquid and the solid phase reported in Appendix, Table.~\ref{tab:density_liquid_solid} were obtained from separate $NpT$ simulations carried out at $T_m$ using 2304 water molecules and simulation times of 1\,ns (after equilibrating for 0.5\,ns). In order to account for possible differences between the DFT and the NNP potential-energy surfaces, the melting temperatures were corrected using thermodynamic perturbation theory as described in the Appendix. All correction terms have positive values which range from 2\,K to 11\,K (Appendix, Table~\ref{tab:melting_point_correction}).

\begin{acknowledgments}
This work was supported by the Cluster of Excellence RESOLV (EXC 1069) funded by the Deutsche Forschungsgemeinschaft as well as by the DFG (Emmy Noether project Be3264/3-1, Heisenberg fellowship Be3264/6-1, and project Be3264/5-1). T.M. is grateful for a PhD fellowship of the Studienstiftung des Deutschen Volkes and for support by the Ruhr-University Research School Plus (DFG GSC 98/3). A.S. is grateful for support by the VSC Research Center funded by the Austrian Federal Ministry of Science, Research and Economy (bmwfw). Financial support of the Austrian Science Fund FWF (Projects P24681-N20 and SFB Vicom, F41) is gratefully acknowledged. The results presented here have been achieved in part using the Vienna Scientific Cluster (VSC). The authors thank S. Imoto, H. Forbert, D. Marx, and M. Heyden for insightful discussions and providing AIMD data and A. Urban and N. Artrith for help with VASP and Wannier90.
\end{acknowledgments}

\cleardoublepage
%

\cleardoublepage

\appendix

\setcounter{figure}{0}
\makeatletter 
\renewcommand{\thefigure}{S\@arabic\c@figure}
\makeatother

\setcounter{table}{0}
\makeatletter 
\renewcommand{\thetable}{S\@arabic\c@table}
\makeatother

\setlength\tabcolsep{5.0pt}

\section{Neural Network Potentials for Bulk Water}

\textit{Ab initio}-quality neural network potentials (NNPs) for water were constructed based on the high-dimensional NNP approach by Behler and Parrinello~\cite{Behler2007}. In this method, the total energy $E$ is written as a sum of atomic energy contributions $E^{\text{hydrogen/oxygen}}$,
\begin{equation}
 E = \sum_{i=1}^{N^{\text{hydrogen}}} E^{\text{hydrogen}}_i + \sum_{j=1}^{N^{\text{oxygen}}} E^{\text{oxygen}}_j,
\end{equation}
which are expressed by artificial neural networks and depend on the local chemical environment represented by a set of atom-centered symmetry functions~\cite{Behler2011}.

High-dimensional NNPs enable constructing highly accurate and full-dimensional representations of reference po\-ten\-ti\-al-energy surfaces for periodic and non-periodic systems~\cite{Behler2014}. While to date this method has been primarily applied to solid state systems~\cite{Behler2014}, recently high-dimensional NNPs have been employed to describe water clusters in the gas phase~\cite{Morawietz2013}, and the interaction between water molecules and bimetallic nanoparticles~\cite{Artrith2014}. The present work represents the first construction of NNPs for a condensed molecular system.

\begin{figure*}[htbp]
\centering
  \includegraphics{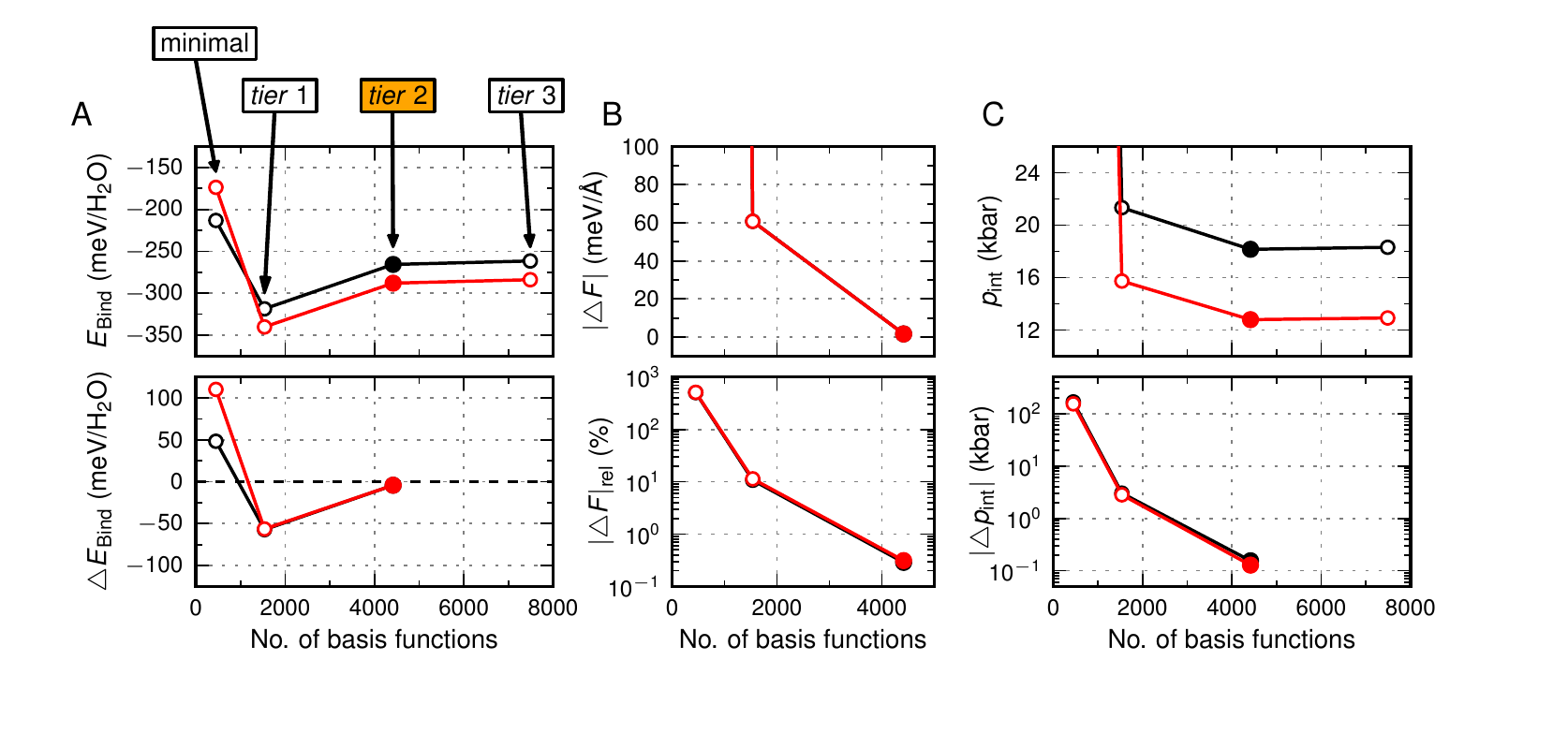}\\  
\caption{\label{fig:basis_set_convergence}\textbf{Basis set convergence.} Convergence of binding energies, $E_{\text{Bind}}$ (A), forces, $F$ (B), and instantaneous pressure, $p_{\text{int}}$ (C) with basis set size for two typical liquid water configurations containing 64 water molecules. $\Delta E_{\text{Bind}} = E_{\text{Bind}} - E_{\text{Bind},\,tier\,3}$, $|\Delta F| = \tfrac{1}{3N_{\text{atoms}}}\sum_{i=1}^{3N_{\text{atoms}}} |F_i - F_{i,\,tier\,3}|$, $|\Delta F|_{\text{rel}} = 100 \times |\Delta F|/\tfrac{1}{3N_{\text{atoms}}}\sum_{i=1}^{3N_{\text{atoms}}} |F_{i,\,tier\,3}|$, $p_{\text{int}} = -\tfrac{1}{3}tr(\text{stress tensor})$, and $|\Delta p_{\text{int}}| = |p_{\text{int}} - p_{\text{int},\,tier\,3}|$. The \textit{tier} 2 basis was chosen as production basis set for all reference calculations used to train the NNPs.}
\end{figure*}

We developed a set of four NNPs trained to energies and forces from reference DFT calculations for a broad range of condensed water configurations employing the RPBE~\cite{Hammer1999} and BLYP~\cite{Becke1988,Lee1988} density-functionals with and without vdW corrections.
All DFT calculations were carried out with the all-electron code FHI-aims~\cite{Blum2009} which uses numerical atom-centered orbitals as basis functions. Since it has been shown that AIMD simulations with underconverged basis sets yield underestimated water densities~\cite{Ma2012}, we carefully checked the convergence of our DFT calculations with respect to basis set size (see Fig.~\ref{fig:basis_set_convergence}). For the chosen ``\textit{tier} 2'' basis set, binding energies, forces, and pressure tensors are well converged, with remaining errors below 4 meV/H$_2$O, 2 meV/\AA{}, and 1.0\,\%, respectively.

VdW interactions were taken into account by Grimme's D3 method~\cite{Grimme2010} using the zero-damping scheme and neglecting three-body contributions since their effect on the properties of water is negligible~\cite{Jonchiere2011}. Within the D3 approach, computing the vdW correction term for periodic structures is computationally more demanding than the evaluation of the NNPs. Therefore, vdW energies and forces were added to the reference data before training the NNPs (in contrast to previously developed NNPs for water clusters~\cite{Morawietz2013} where vdW interactions were computed on-the-fly in the application of the potentials).

The NNPs were trained using the code RuNNer~\cite{Behler2009}. All NNPs consist of a set of feed-forward neural networks with two hidden layers containing 25 nodes each, corresponding to a total number of 2,827 parameters per NNP. For the nodes in the hidden layers the hyperbolic tangent was used as activation function, while for the node in the output layer a linear function was used. Local chemical environments were described by symmetry functions of type 2 and type 4~\cite{Behler2011}. The radial extension of each symmetry function is restricted by a cutoff function $f_c$ that ensures that value and slope become zero at the cutoff radius $r_c$,
\begin{align}
f_c(r_{ij})&=\begin{cases} \tanh^3 \Bigl[1 - \frac{r_{ij}}{r_c}\Bigr] & \text{with}\quad r_{ij}\leq r_c  \\  0 & \text{with}\quad r_{ij}> r_c. \end{cases}
\end{align}
Here, $r_{ij}$ is the distance between atoms $i$ and $j$. Radial symmetry functions of type 2 consist of a sum of distance dependent Gaussian functions multiplied by the cutoff function,
\begin{align}
 G^2_i=\sum_{j\neq i} e^{-\eta(r_{ij}-r_s)^2} f_c(r_{ij}).
\end{align}
Here, the center of the Gaussian can be shifted by the parameter $r_s$ and the parameter $\eta$ can be adjusted in order to change the Gaussian width. The angular symmetry function of type 4 consists of an angular term that depends on the angle $\alpha_{ijk}$ (centered at atom $i$ and formed with neighbors $j$ and $k$) and can be adjusted by varying the parameters $\lambda$ and $\zeta$. Additional terms that depend on the interatomic distances of atoms $i$, $j$, and $k$ control the radial resolution,
\begin{align}
 G^4_i=&2^{1-\zeta}\sum_{j\neq i}\sum_{k\neq i,j} \Biggl[\bigl(1+\lambda cos(\alpha_{ijk})\bigl)^\zeta \\
 &\times e^{-\eta(r^2_{ij}+r^2_{ik}+r^2_{jk})} f_c(r_{ij}) f_c(r_{ik}) f_c(r_{jk}) \Biggr].  \nonumber
\end{align}
A total of 27 and 30 symmetry functions were used to describe the atomic environments of hydrogen and oxygen atoms, respectively. The parameters of the symmetry functions are listed in Tables~\ref{tab:symm_functions_hydrogen} and \ref{tab:symm_functions_oxygen}.

\begin{table}[htpb]
\centering
\caption{\label{tab:symm_functions_hydrogen} \textbf{Symmetry function parameters for hydrogen.} Parameters $r_s$ (in Bohr), $\eta$ (in Bohr$^{-2}$), $\lambda$, and $\zeta$ of atom-centered symmetry functions of type $G^2$ (radial, nos. 1 -- 16) and type $G^4$ (angular, nos. 17 -- 27) used to describe the local chemical environments of hydrogen atoms. The cutoff radius $r_c$ is 12\,Bohr ($\approx$~6.35\,\AA{}) for all symmetry functions.
}
\begin{tabular}{ccccrrl}
 No. &  Element $j$ & Element $k$ & $r_s$ & $\eta$ & $\lambda$ & $\zeta$ \\
 \hline\noalign{\smallskip}
 \noalign{\smallskip}
    1&   H& --- &   0.0&               0.001 & --- & --- \\
    2&   O& --- &   0.0&               0.001 & --- & --- \\
    3&   H& --- &   0.0&               0.010 & --- & --- \\    
    4&   O& --- &   0.0&               0.010 & --- & --- \\    
    5&   H& --- &   0.0&               0.030 & --- & --- \\
    6&   O& --- &   0.0&               0.030 & --- & --- \\
    7&   H& --- &   0.0&               0.060 & --- & --- \\   
    8&   O& --- &   0.0&               0.060 & --- & --- \\   
    9&   O& --- &   0.9&               0.150 & --- & --- \\
   10&   H& --- &   1.9&               0.150 & --- & --- \\    
   11&   O& --- &   0.9&               0.300 & --- & --- \\   
   12&   H& --- &   1.9&               0.300 & --- & --- \\   
   13&   O& --- &   0.9&               0.600 & --- & --- \\
   14&   H& --- &   1.9&               0.600 & --- & --- \\
   15&   O& --- &   0.9&               1.500 & --- & --- \\
   16&   H& --- &   1.9&               1.500 & --- & --- \\
17&  O & O & 0.0& 0.001 &   -1.0 &  4.0  \\
18&  O & O & 0.0& 0.001 &    1.0 &  4.0  \\
19&  H & O & 0.0& 0.010 &   -1.0 &  4.0  \\
20&  H & O & 0.0& 0.010 &    1.0 &  4.0  \\
21&  H & O & 0.0& 0.030 &   -1.0 &  1.0  \\
22&  O & O & 0.0& 0.030 &   -1.0 &  1.0  \\
23&  H & O & 0.0& 0.030 &    1.0 &  1.0  \\
24&  O & O & 0.0& 0.030 &    1.0 &  1.0  \\
25&  H & O & 0.0& 0.070 &   -1.0 &  1.0  \\
26&  H & O & 0.0& 0.070 &    1.0 &  1.0  \\
27&  H & O & 0.0& 0.200 &    1.0 &  1.0  \\
 \hline
\end{tabular}
\end{table}

\begin{table}[htpb]
\centering
\caption{\label{tab:symm_functions_oxygen} \textbf{Symmetry function parameters for oxygen.} Parameters $r_s$ (in Bohr), $\eta$ (in Bohr$^{-2}$), $\lambda$, and $\zeta$ of atom-centered symmetry functions of type $G^2$ (radial, nos. 1 -- 16) and type $G^4$ (angular, nos. 17 -- 30) used to describe the local chemical environments of oxygen atoms. The cutoff radius $r_c$ is 12\,Bohr ($\approx$~6.35\,\AA{}) for all symmetry functions.
}
\begin{tabular}{ccccrrl}
 No. &  Element $j$ & Element $k$ & $r_s$ & $\eta$ & $\lambda$ & $\zeta$ \\
 \hline\noalign{\smallskip}
 \noalign{\smallskip}
    1&  H& --- &   0.0&               0.001 & --- & ---  \\
    2&  O& --- &   0.0&               0.001 & --- & ---  \\
    3&  H& --- &   0.0&               0.010 & --- & ---  \\
    4&  O& --- &   0.0&               0.010 & --- & ---  \\
    5&  H& --- &   0.0&               0.030 & --- & ---  \\
    6&  O& --- &   0.0&               0.030 & --- & ---  \\
    7&  H& --- &   0.0&               0.060 & --- & ---  \\
    8&  O& --- &   0.0&               0.060 & --- & ---  \\
    9&  H& --- &   0.9&               0.150 & --- & ---  \\
   10&  O& --- &   4.0&               0.150 & --- & ---  \\
   11&  H& --- &   0.9&               0.300 & --- & ---  \\    
   12&  O& --- &   4.0&               0.300 & --- & ---  \\    
   13&  H& --- &   0.9&               0.600 & --- & ---  \\
   14&  O& --- &   4.0&               0.600 & --- & ---  \\
   15&  H& --- &   0.9&               1.500 & --- & ---  \\
   16&  O& --- &   4.0&               1.500 & --- & ---  \\
17&  H & O & 0.0&  0.001 &   -1.0 &  4.0  \\
18&  O & O & 0.0&  0.001 &   -1.0 &  4.0  \\
19&  H & O & 0.0&  0.001 &    1.0 &  4.0  \\
20&  O & O & 0.0&  0.001 &    1.0 &  4.0  \\
21&  H & H & 0.0&  0.010 &   -1.0 &  4.0  \\
22&  H & H & 0.0&  0.010 &    1.0 &  4.0  \\
23&  H & H & 0.0&  0.030 &   -1.0 &  1.0  \\
24&  H & O & 0.0&  0.030 &   -1.0 &  1.0  \\
25&  O & O & 0.0&  0.030 &   -1.0 &  1.0  \\
26&  H & H & 0.0&  0.030 &    1.0 &  1.0  \\
27&  H & O & 0.0&  0.030 &    1.0 &  1.0  \\
28&  O & O & 0.0&  0.030 &    1.0 &  1.0  \\
29&  H & H & 0.0&  0.070 &   -1.0 &  1.0  \\
30&  H & H & 0.0&  0.070 &    1.0 &  1.0  \\
 \hline
\end{tabular}
\end{table}

The functional form of the atomic neural networks describing hydrogen and oxygen atoms is then given by,
\begin{align}
E^{\text{hydrogen}} =& f^3_1\biggl(b^3_1+\sum^{25}_{k=1}a^{23}_{k1}f^2_k\biggl(b^2_k+\sum^{25}_{j=1}a^{12}_{jk}\\\nonumber
&\times f^1_j\biggl(b^1_j+\sum^{27}_{i=1}a^{01}_{ij}G_i\biggr)\biggr)\biggr),
\end{align}
and,
\begin{align}
E^{\text{oxygen}} =& f^3_1\biggl(b^3_1+\sum^{25}_{k=1}a^{23}_{k1}f^2_k\biggl(b^2_k+\sum^{25}_{j=1}a^{12}_{jk}\\\nonumber
&\times f^1_j\biggl(b^1_j+\sum^{30}_{i=1}a^{01}_{ij}G_i\biggr)\biggr)\biggr),
\end{align}
respectively. Here, the weight parameters $a^{kl}_{ij}$ together with the bias weights $b_j^i$ are the fitting parameters of the NNP and $f^1$, $f^2$, and $f^3$ are activation functions with the following functional form:
\begin{align}
 f^1(x) =& \text{tanh}(x), \\
 f^2(x) =& \text{tanh}(x), \\
 f^3(x) =& x.
\end{align}
The force component $F_{\alpha_k}$ acting on atom $k$ in direction $\alpha$\,\,$=x,\,\,y,\,\text{or}\,\,\,z$, given by the negative gradient of the energy with respect to $\alpha_k$, is obtained from,
\begin{align}
 F_{\alpha_k} = -\frac{\partial E}{\partial\alpha_k} =-\sum^N_{i=1}\frac{\partial E_i}{\partial\alpha_k} = -\sum^N_{i=1}\sum^{M_i}_{j=1}\frac{\partial E_i}{\partial G_{i,j}}\frac{\partial G_{i,j}}{\partial\alpha_k},
\end{align}
where the derivatives $\partial E_i/\partial G_{i,j}$ and $\partial G_{i,j}/\partial\alpha_k$ are defined by the functional form of the atomic neural networks and the symmetry functions, respectively. Here, $N$ is the number of atoms and $M_i$ the number of symmetry functions of atom $i$.

In order to avoid a saturation of the activation functions in the first hidden layer, the initial symmetry function values $G_i^0$ are always centered and rescaled,
\begin{align}
 G_i = \frac{G^0_i - G^0_{i,\text{average}}}{G^0_{i,\text{max}} - G^0_{i,\text{min}}},
\end{align}
using the average, maximum, and minimum symmetry function values obtained from the full reference data set. The derivatives are modified correspondingly:
\begin{align}
 \frac{\partial G_i}{\partial\alpha_k} = \frac{\partial G_i^0}{\partial\alpha_k} \frac{1}{G^0_{i,\text{max}} - G^0_{i,\text{min}}}.
\end{align}
Parameter sets (weight parameter and bias weights) as well as average, minimum, and maximum values for each symmetry function are available online for all NNPs~\cite{figshare2015}.

Comparisons of potentials with and without explicit consideration of long-range electrostatics~\cite{Artrith2011,Morawietz2012} have shown that there is no significant difference in the accuracy of the energies and forces in the training and in the test set for the chosen cutoff radius of 6.35\,\AA{}. Consequently, long-range electrostatics were not included explicitly and the use of Ewald summation techniques is avoided, ensuring a linear scaling of the computational costs with system size.

\begin{figure*}[htbp]
\centering
  \includegraphics{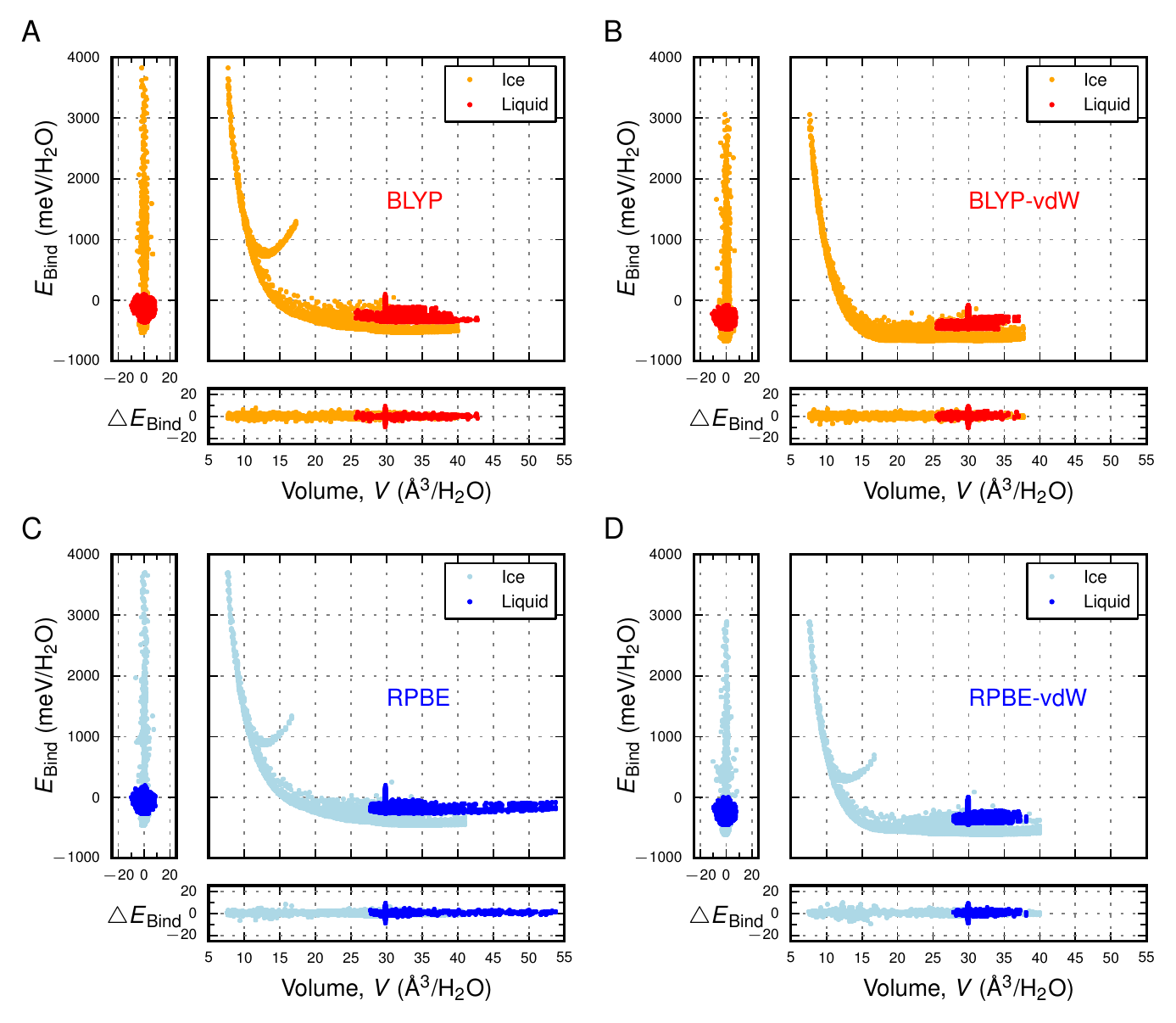}\\  
\caption{\label{fig:reference_configs} \textbf{Accuracy and distribution of reference configurations.} Energy error $\Delta E_{\text{Bind}} = E_{\text{Bind, NNP}} -E_{\text{Bind, DFT}}$ (in meV/H$_2$O) as function of binding energy and volume, and location of reference configurations in the $E_{\text{Bind}}$ vs. $V$ space for NNPs based on BLYP (A), BLYP-vdW (B), RPBE (C), and RPBE-vdW (D).}
\end{figure*}

Individual data sets for each reference method were generated in an iterative process starting with initial data sets that were systematically extended. Reference data sets were always randomly split into a training set, containing 90\% of all configurations, and an independent test set, containing the remaining 10\% of configurations. Initial reference configurations contain crystalline configurations obtained from DFT relaxations and liquid configurations from force field MD simulations. Configurations of eight different ice polymorphs (ice I\textit{h}, XI, IX, II, XIV, XV, VIII, and X) at different lattice constants were included. In addition to the fully relaxed configurations, also distorted structures with randomly displaced atomic positions were used. Initial configurations for liquid water were taken from force field MD simulations at different temperatures, employing the simulation package GROMACS~\cite{VanderSpoel2005} and the flexible non-polarizable SPC/Fw~\cite{Wu2006} water model, and recomputed with the respective reference method. Simulations with 16 and 32 water molecules were performed at the experimental density of water, while unit cells containing 64 water molecules were employed for \textit{NpT} simulations at various densities. Based on these data, preliminary NNPs were constructed and employed in structural relaxations and MD simulations (with units cells containing up to 128 water molecules) at various temperatures and pressures to generate new configurations, which were then recomputed by DFT and added to the initial data sets. After four cycles of refinement the NNPs were converged and applied in production runs. Final reference data sets contain about 7,000 periodic configurations per NNP, corresponding to $\approx$~1,700,000 force components, which have also been used for training the NNPs. As illustrated in Fig.~\ref{fig:reference_configs}, the energy error does not grow with increasing binding energy, and all configurations, independent of their location in the energy vs. volume phase space, are equally well represented.

\begin{figure}[htbp]
 \centering
  \includegraphics{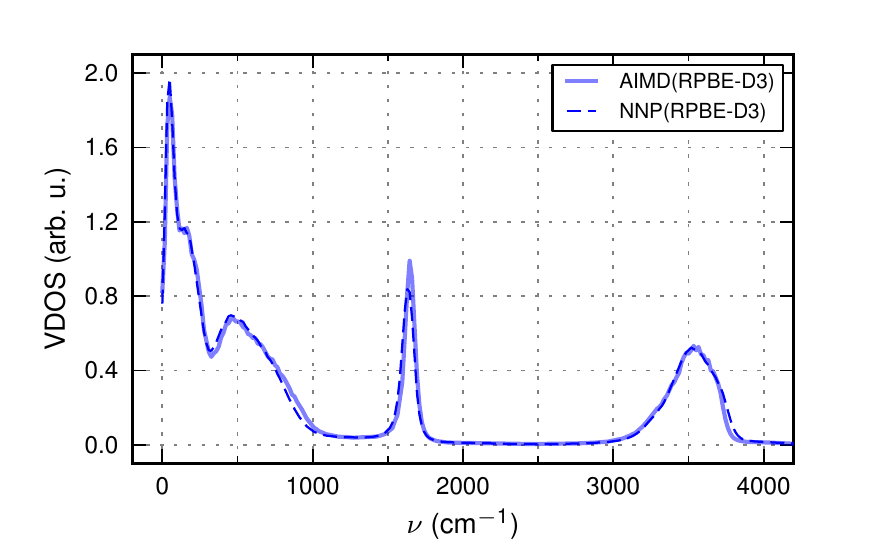}\\  
 \caption{\label{fig:VDOS}\textbf{VDOS spectrum.} Vibrational density of state (VDOS) computed from velocity autocorrelation functions obtained from RPBE-D3-based NNP simulations at $T$\,=\,$300$\,K and $\rho =1$\,g/cm$^{3}$ compared to results from \textit{ab initio} simulations~\cite{Imoto2015}.
}
\end{figure}

\section{Viscosities and Diffusion Coefficients}

\begin{figure*}[htbp]
\centering
  \includegraphics{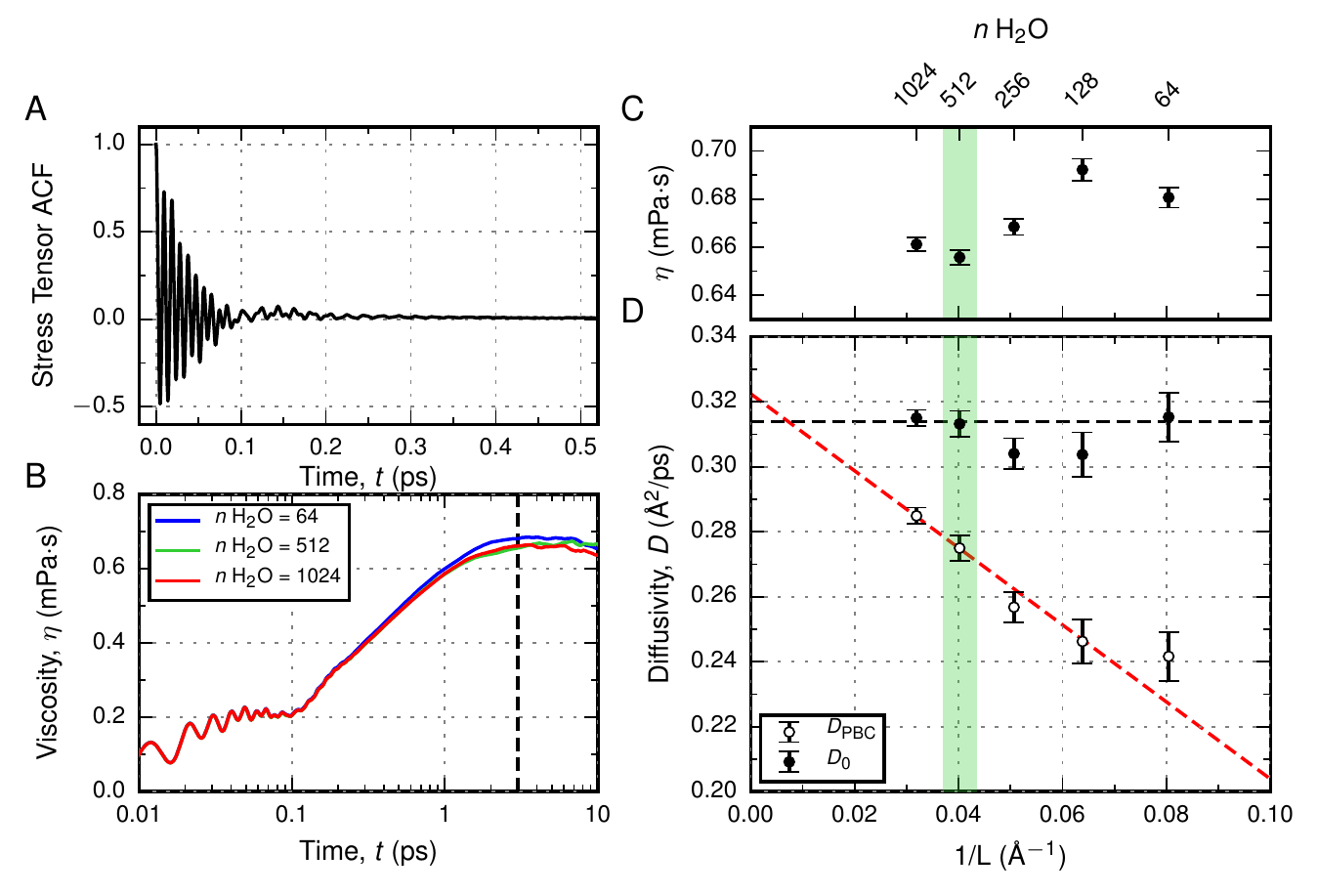}\\  
\caption{\label{fig:diffusion_viscosity} 
\textbf{System-size dependence of diffusion coefficient and viscosity.}
(A) Normalized stress autocorrelation function for a system of 64 water molecules.
(B) Running values of the viscosity $\eta$ for different system sizes computed from the stress autocorrelation function using the Green-Kubo formula. The dashed line indicates the upper limit used for computing the final viscosity values.
(C-D) Viscosity, and uncorrected $D_{\text{PBC}}$ and corrected $D_0$ diffusion coefficients as function of inverse box length 1/$L$. The red and black dashed lines are weighted least squares fits of the uncorrected and corrected diffusion coefficients, respectively.
All values reported here are obtained from NNP simulations based on the vdW-corrected RPBE density-functional performed at $T$\,=\,$300$\,K and $\rho =1$\,g/cm$^{3}$. The green bar indicates the system size used for computing the results reported in Fig.~\ref{fig:fig01} of the main text.
}
\end{figure*}

Shear viscosities $\eta$ were computed from the Green-Kubo relation,
\begin{align}
 \eta = \frac{V}{k_{\text{B}}T}\int_0^\infty\bigl< P_{\alpha\beta}(t)P_{\alpha\beta}(0)\bigr> dt,
\end{align}
where $\bigl< P_{\alpha\beta}(t)P_{\alpha\beta}(0)\bigr>$ is the autocorrelation function of the stress tensor element $P_{\alpha\beta}$. Autocorrelation functions (cf. Fig.~\ref{fig:diffusion_viscosity}a) were averaged over the five independent components $P_{xy}$, $P_{xz}$, $P_{yz}$, $\frac{1}{2}(P_{xx}-P_{yy})$, and $\frac{1}{2}(P_{yy}-P_{zz})$. A value of 3\,ps was chosen for the upper limit of the integral (see Fig.~\ref{fig:diffusion_viscosity}b). As shown in Fig.~\ref{fig:diffusion_viscosity}c, the final viscosity values are essentially system size independent.

Diffusion coefficients $D_{\text{PBC}}$ were computed from mean squared displacements,
\begin{align}
 D_{\text{PBC}} = \lim_{t\to\infty} \frac{1}{6}\frac{d}{dt}\bigl<\bigl| \mathbf{r}(t) - \mathbf{r}(0) \bigl|^2\bigr>.
\end{align}
In order to correct for finite size effects, we obtained system size independent diffusion coefficients $D_0$ (cf. Fig.~\ref{fig:diffusion_viscosity}d) from the relation~\cite{Dunweg1993,Yeh2004},
\begin{align}
 D_0 = D_{\text{PBC}} + \frac{k_{\text{B}}T\xi}{6\pi}\frac{1}{\eta L},
\end{align}
where $L$ is the length of the cubic simulation cell and $\xi= 2.837297$. The viscosity values and diffusion coefficients shown in Fig.~\ref{fig:fig01} in the main text were computed for 512 water molecules. Error bars were estimated from the standard error of the mean of the values obtained from the 32 independent \textit{NVE} trajectories. Due to the very low water mobility in the BLYP simulations, converged viscosity values could not be computed for $T = 275$ and $300$\,K.

\section{Hydrogen Bond Analysis}

The hydrogen bond (HB) autocorrelation functions $c(t)$ and $n(t)$, shown in Fig.~\ref{fig:fig01} in the main text, were obtained employing the HB criterion by Luzar and Chandler~\cite{Luzar1996a} (wherein a pair of water molecules is considered hydrogen bonded if r$_{\text{OO}}<3.5$\,\AA{} and $\beta=\angle\text{H}_\text{D}\!\!-\!\text{O}_\text{D}\!\cdots\!\text{O}_\text{A}<30^{\circ}$). The autocorrelation function $c(t)$ is given by,
\begin{align}
c(t) = \frac{\bigl<h(t)h(0)\bigr>}{h},
\end{align}
where $h(t)$ is unity if a particular pair of water molecules is hydrogen bonded at time $t$ and is zero otherwise~\cite{Luzar1996}. $c(t)$ is the intermittent HB autocorrelation function which does not require that a particular HB remains continuously intact but also counts HBs that break and subsequently reform. The autocorrelation function $n(t)$ gives the time-dependent probability that a water pair that is \textit{not} hydrogen bonded remains within a distance of 3.5\,\AA{} from each other and is defined by,
\begin{align}
n(t) = \frac{\bigl<h(0)[1-h(t)]H(t)\bigr>}{h},
\end{align}
where $H(t)$ is set to unity if the water pair is closer than 3.5\,\AA{} and is zero otherwise. In addition to the correlation functions, forward and backward rate constants and HB relaxation times and lifetimes were computed based on the Luzar-Chandler model~\cite{Luzar1996} and are reported in Table~\ref{tab:hbond_kinetics}.

\begin{table*}[htbp]
\centering
\caption{\label{tab:hbond_kinetics} \textbf{Hydrogen bond kinetics.} Comparison of hydrogen bond relaxation time ($\tau_{\text{rlx}}$), forward and backward rate constants (k and k'), and lifetime ($\tau_{\text{HB}}$) based on the Luzar-Chandler model~\cite{Luzar1996}. In addition, the average number of hydrogen bonds ($n$ HB) is shown. The NNP values were obtained from simulations at \textit{T}\,=\,300\,K. The TIP4P results were taken from Ref.~\cite{Xu2002}.
}
\begin{tabular}{lllllll}
Model & \multicolumn{1}{c}{$\tau_{\text{rlx}}$ (ps)} & \multicolumn{1}{c}{k (ps$^{-1}$)} & \multicolumn{1}{c}{k' (ps$^{-1}$)} & \multicolumn{1}{c}{$\tau_{\text{HB}}$ (ps)} & \multicolumn{1}{c}{$\tau_{\text{rlx}}$/$\tau_{\text{HB}}$} & \multicolumn{1}{c}{$n$ HB} \\
 \hline\noalign{\smallskip}
NNP(BLYP)	& --- & --- & --- & --- & --- & 3.81 \\
NNP(BLYP-vdW)	& 7.12 & 0.24 &	0.57 & 4.22 & 1.69 & 3.64 \\
NNP(RPBE)	& 2.00 & 1.15 &	8.37 & 0.87 & 2.30 & 2.61 \\
NNP(RPBE-vdW)	& 4.33 & 0.45 &	1.65 & 2.24 & 1.93 & 3.47 \\
 \noalign{\smallskip}
TIP4P	& 3.32 & 0.45 &	1.02 & 2.22 & 1.49 & 3.54 \\
 \hline\noalign{\smallskip}
\end{tabular}	
\end{table*}

\section{Properties of Crystalline Water}

\begin{figure*}[htbp]
 \centering
  \includegraphics{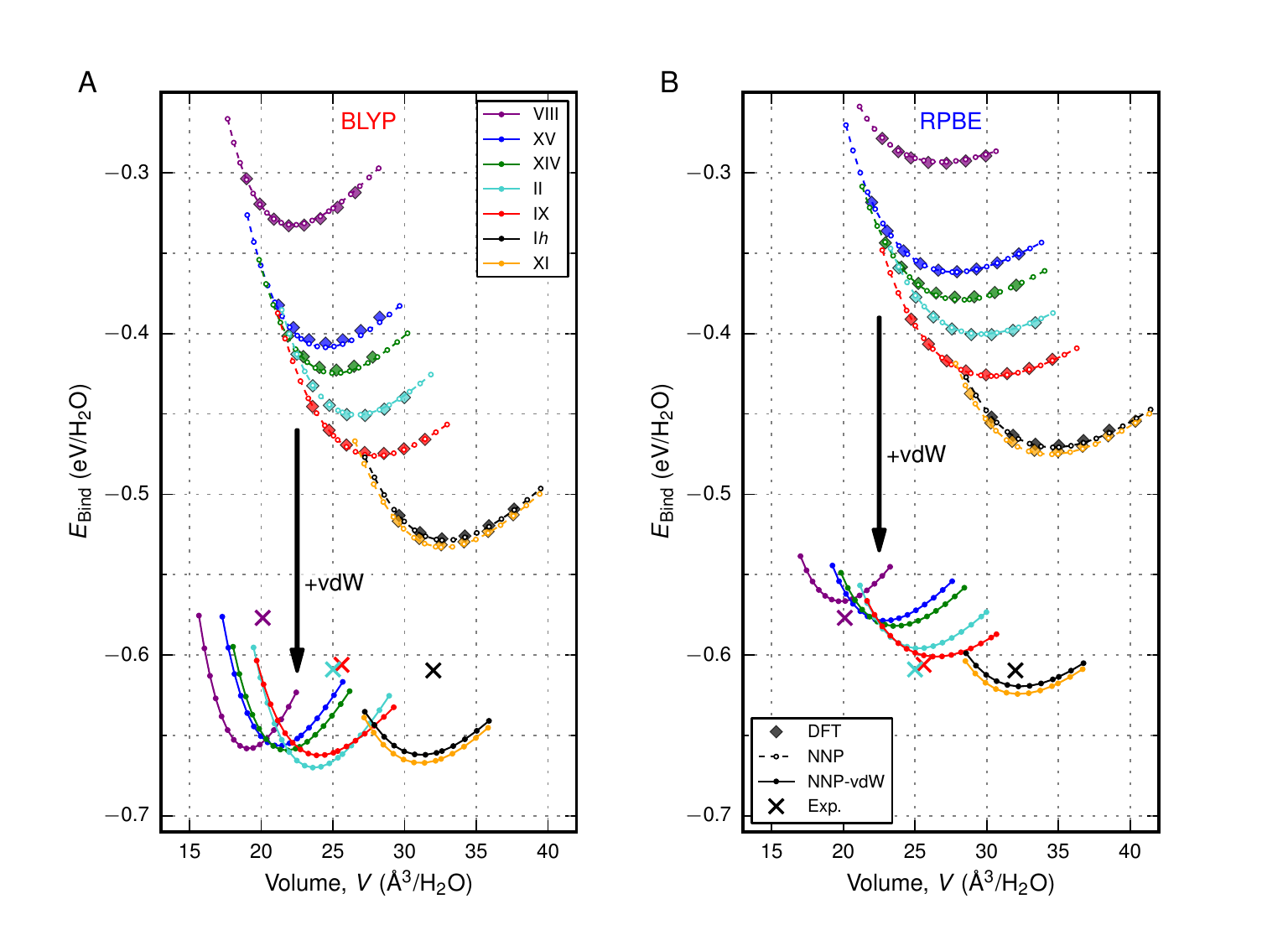}\\  
 \caption{\label{fig:ice_e_vs_v} \textbf{Energy vs. volume curves.} Lattice energy (\textit{T} = 0\,K) as function of volume of seven ice phases computed with DFT and NNPs based on BLYP / BLYP-vdW (A) and RPBE / RPBE-vdW (B). For ice I\textit{h}, IX, II, and VIII also the experimental equilibrium values~\cite{Rottger1994,Londono1993,Whalley1984} are shown.}
\end{figure*}

Structural and energetical properties of seven low- to high-pressure ice polymorphs (ice I\textit{h}, XI, IX, II, XIV, XV, and VIII) were computed using the NNPs and DFT and compared to the corresponding experimental values~\cite{Rottger1994,Line1996,Londono1993,Whalley1984,Salzmann2006,Salzmann2009}. Experimental lattice energies (taken from Ref.~\cite{Whalley1984}) are extrapolated to 0 K and do not contain zero-point contributions. Energy vs. volume curves were computed by isotropic variation~\cite{Santra2013a} of the experimental lattice parameters followed by a full relaxation of all atoms in the unit cell employing the L-BFGS algorithm~\cite{Liu1989}. As shown in Fig.~\ref{fig:ice_e_vs_v}, curves obtained from NNP calculations closely reproduce the reference DFT values and the inclusion of vdW interactions leads to a significantly improved agreement with experiment. Equilibrium lattice energies and volumes were obtained by fitting the Murnaghan equation of state~\cite{Murnaghan1944} to the energy vs. volume curves. Deviations of the equilibrium values between NNP and DFT are only a small fraction of the errors of the DFT values with respect to experiment.

\section{Dielectric Properties}

\begin{figure*}[htbp]
\centering
  \includegraphics{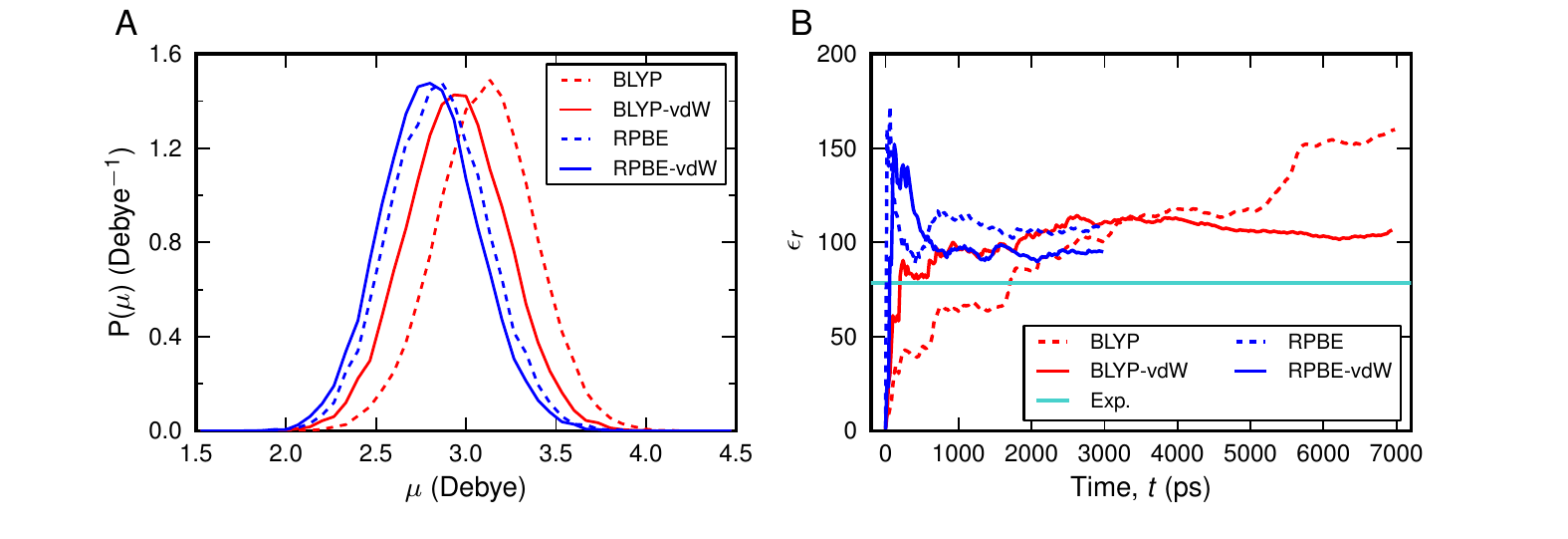}\\  
\caption{\label{fig:dielectrics}\textbf{Dielectric properties.}
(A) Probability density function P($\mu$) of the molecular dipole moment magnitude $\mu$ computed from maximally localized Wannier functions evaluated at configurations sampled by NNPs at 300\,K. (B) Convergence of the cumulative average of the dielectric constant $\epsilon_r$ as function of simulation time.
}
\end{figure*}

Molecular dipole moments $\boldsymbol{\mu}$ and dielectric constants $\epsilon_r$ were obtained by post-processing independent configurations from NNP simulations in the $NVT$ ensemble at $T = 300$\,K employing unit cells containing 128 molecules at the experimental density ($\rho = 0.9965$\,g/cm$^3$). After equilibrating for 1\,ns, simulations were carried out for 3\,ns and 7\,ns for the RPBE- and BLYP-based potentials, respectively. Due to their reduced water dynamics (cf. Fig.~\ref{fig:fig01} in the main text), longer simulation times were employed for the BLYP-based NNPs. Configurations were extracted every 20\,ps and maximally localized Wannier functions~\cite{Marzari1997} (MLWFs) were computed using the projector augmented wave~\cite{Blochl1994,Kresse1999} (PAW)-based Vienna \textit{ab initio} simulation package~\cite{Kresse1993,Kresse1994,Kresse1996,Kresse1996a} (VASP, employing a plane-wave cutoff of 700 eV) and the WANNIER90 program~\cite{Mostofi2008}. Molecular dipole moments $\boldsymbol{\mu}$ were computed using the Wannier function centers (WFCs) of the four MLWFs representing the valence electrons. The dielectric constant $\epsilon_r$ was calculated using the the relation,
\begin{align}
\epsilon_r = \frac{1}{3k_{\text{B}}\epsilon_0TV}\left(\left<\mathbf{M}^2\right> - \left<\mathbf{M}\right>^2\right) + \epsilon_\infty,
\end{align}
where $T$ is the temperature, $V$ is the box volume, $\mathbf{M} = \sum_i^N \mu_i$ is the total dipole moment of the simulation box and $\epsilon_\infty$ is the permittivity of vacuum ($\epsilon_\infty$ = 1.8~\cite{Lu2008}). Fig.~\ref{fig:dielectrics} shows the distribution of the molecular dipole moment and the convergence of $\epsilon_r$ with simulation time for all NNPs. The average magnitudes of the molecular dipole moment and the final values for the dielectric constant are reported in Table~\ref{tab:dielectrics}.

\begin{table}
\centering
\caption{\label{tab:dielectrics}\textbf{Dielectric properties.} Dielectric constant $\epsilon_r$ and average magnitude of the molecular dipole moment $\mu$ from NNP simulations at 300\,K.}
\begin{tabular}{lrl}
Model & \multicolumn{1}{c}{$\epsilon_r$} & \multicolumn{1}{c}{$\mu$ (D)} \\
\hline\noalign{\smallskip}
NNP(BLYP)                               &       159.8          	&      	3.11             \\
NNP(BLYP-vdW)                           &       106.2        	&      	2.95	         \\
NNP(RPBE)                               &       108.9          	&      	2.85             \\
NNP(RPBE-vdW)                           &       95.2          	&      	2.80	         \\
\hline
\end{tabular}
\end{table}

\section{Density Isobars}

\begin{figure}[htbp]
 \centering
  \includegraphics{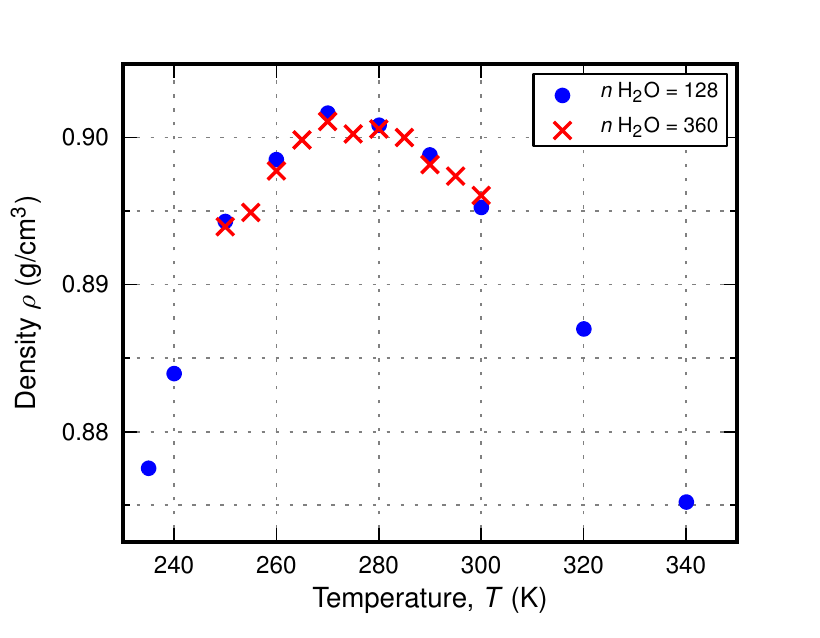}\\  
 \caption{\label{fig:system_size_density_isobar} \textbf{System size dependence of the density isobar.} Density isobars at \textit{p}\,=\,1\,bar obtained from NNP(RPBE-vdW) simulations employing simulation cells containing 128 (blue filled circles) and 360 (red crosses) water molecules, respectively. 
}
\end{figure}

In order to evaluate the influence of system size on the computed density isobars, we performed additional NNP simulations at temperatures around the density maximum for a larger system ($n$\,H$_2$O = 360) which are compared to the results obtained for the smaller system ($n$\,H$_2$O = 128). As shown in Fig.~\ref{fig:system_size_density_isobar}, no significant difference between the two density isobars is visible, indicating that the results obtained for 128 molecules are converged with respect to system size.

Values for the temperature of maximum density, the density at the maximum, and the thermal expansivity, $\alpha$,
\begin{align}
 \alpha = - \frac{1}{\rho}\biggl(\frac{\partial\rho}{\partial T}\biggr)_p,
\end{align}
at ambient conditions obtained for simulations cells containing 128 molecules are reported in Table~\ref{tab:density_max}.

\begin{table}[htbp]
\centering
\caption{\label{tab:density_max}\textbf{Density maximum and thermal expansivity.} Comparison of temperature of maximum density (TMD) (in K), density at \textit{T}\,=\,TMD ($\rho_{\text{TMD}}$) (in g/cm$^3$), and coefficient of thermal expansion at \textit{T}\,=\,25\,$^{\circ}$C ($\alpha_{\text{25\,$^{\circ}$C}}$) (in 10$^{-6}$/K) obtained from \textit{NpT} simulations of 128 H$_2$O using different NNPs. Experimental values were taken from Ref.~\cite{Kell1975}.
}
\begin{tabular}{llll}
Model & \multicolumn{1}{c}{TMD} & \multicolumn{1}{c}{$\rho_{\text{TMD}}$} & \multicolumn{1}{c}{$\alpha_{\text{25\,$^{\circ}$C}}$} \\
 \hline\noalign{\smallskip}
NNP(BLYP)	& {---}		& {---}		& 991	\\
NNP(BLYP-vdW)	& 256		& 1.054		& 435	\\
NNP(RPBE)	& {---}		& {---}		& 2369	\\
NNP(RPBE-vdW)	& 274		& 0.901		& 370	\\
Exp.		& 277.13	& 0.99997	& 257.12\\
 \hline\noalign{\smallskip}
\end{tabular}
\end{table}

\section{Melting Temperatures}

Melting temperatures of ice I\textit{h} were computed employing the interface pinning method~\cite{Pedersen2013,Pedersen2013a}. The method is based on stabilizing a liquid-solid interface in an elongated simulation box (see Fig.~\ref{fig:fig02}d in the main text) by pinning it with an order parameter-dependent bias energy $V_B(R)=\frac{\kappa}{2} \left[Q(\mathbf{R}) - a \right]^2$, where $\kappa$ is a spring constant and $a$ is the anchor point. The order parameter $Q(\mathbf{R})$ quantifies the crystalline fraction of the system. The difference $\Delta \mu$ in chemical potential between the solid and the liquid phase follows from the average deviation of the order parameter from the anchor point. The melting temperature is then determined via an iterative scheme based on the Newton-Raphson method described in Ref.~\cite{Pedersen2013a}. The order parameter was chosen~\cite{Pedersen2013a} as $Q(\mathbf{R}) = \left| \rho_\mathbf{k} \right|$ where $\mathbf{k} = ( 2 \pi n_x / X,\, 2 \pi n_y / Y,\, 0 )$ and $n_x = 6, n_y = 12$. The spring constant and anchor point of the interface pinning term were set to $\kappa =$ \,0.05\,eV and $a =$ \,19.0. All interface pinning simulations were performed using 2304 water molecules, a time step of 1\,fs and a pressure of $p =$ \,1\,bar. The total simulation time was 15\,ns.
The melting temperatures and corresponding errors were derived as follows: from interface pinning simulations we extract pairs of $(T, \Delta \mu \pm \sigma_{\Delta \mu})$, where $\Delta \mu = \mu_\text{liquid} - \mu_\text{solid} = \frac{\kappa \Delta Q}{N} \left[ \left<Q\right>' - a \right]$ and $\sigma_{\Delta \mu}$ is determined from block averages. In the vicinity of $T_m$ we assume a linear dependency $\Delta \mu(T) = k T + d$ and use the pairs $(T, \Delta \mu \pm \sigma_{\Delta \mu})$ to fit $k$ and $d$. We derive $T_m$ via $\Delta \mu(T_m) = 0$ and the errors using 68\% confidence interval bands.

\section{Melting Point Correction}

Due to small differences between the NNP and the DFT energies, the melting temperature $T_m^{\text{NNP}}$ obtained with the NNP may differ from the melting temperature $T_m^{\text{DFT}}$ of the reference method. Using thermodynamic perturbation theory, we next derive a correction term,
\begin{align}
\Delta T_m  =  T_m^{\text{DFT}} - T_m^{\text{NNP}},
\end{align}
which accounts for this difference.

\begin{figure*}[htbp]
 \centering
  \includegraphics{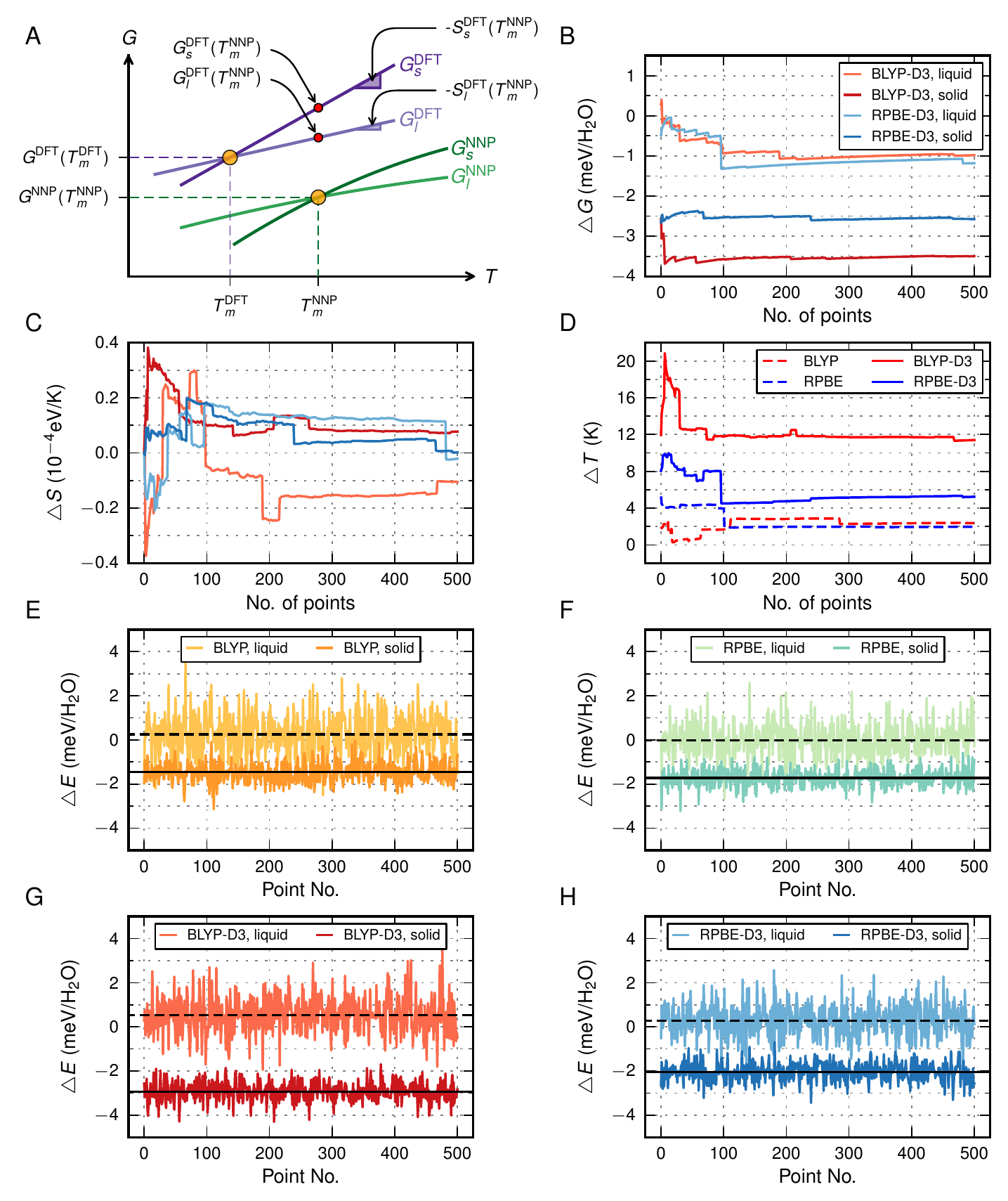}\\  
 \caption{\label{fig:melting_point_correction}
 \textbf{Melting Point Correction.} (A) The correction term $\Delta T_m = T_m^{\text{DFT}} - T_m^{\text{NNP}}$ is estimated by expanding the DFT Gibbs free energy $G^{\text{DFT}}$ as a Taylor series around $T = T_m^{\text{NNP}}$. The slope of the linear expansion is given by the negative entropy of fusion -$S^{\text{DFT}}$. (B-C) Convergence of free energy difference $\Delta G = G^{\text{DFT}} - G^{\text{NNP}}$ and entropy difference $\Delta S = S^{\text{DFT}} - S^{\text{NNP}}$ as function of the number of configurations used to obtain the average quantifies given in Eq.~\eqref{eq:eq04} for the vdW-corrected NNPs. (D) Convergence of the melting point correction term $\Delta T$ for all NNPs. (E-H) Energy difference $\Delta E = E^{\text{DFT}} - E^{\text{NNP}}$ for all configurations used for the melting point correction. The black solid and black dashed lines indicate the average energy difference for the solid and the liquid phase, respectively.
}
\end{figure*}

We first approximate the DFT Gibbs free energy of the liquid phase $G_l^{\text{DFT}}$($p,T$) and the solid phase $G_s^{\text{DFT}}$($p,T$) by a Taylor expansion at $T = T_m^{\text{NNP}}$, truncated after the linear term (see Fig.~\ref{fig:melting_point_correction}a, $p$ omitted for clarity),
\begin{align}
G_l^{\text{DFT}}(T) \approx G_l^{\text{DFT}}(T_m^{\text{NNP}}) + \left. \frac{\partial G_l^{\text{DFT}}}{\partial T}\right |_{T_m^{\text{NNP}}}\bigl(T-T_m^{\text{NNP}}\bigl),\nonumber
\\[8pt]
G_s^{\text{DFT}}(T) \approx G_s^{\text{DFT}}(T_m^{\text{NNP}}) + \left. \frac{\partial G_s^{\text{DFT}}}{\partial T}\right |_{T_m^{\text{NNP}}}\bigl(T-T_m^{\text{NNP}}\bigl).\nonumber
\end{align}
Using the equivalence $G_l^{\text{DFT}} = G_s^{\text{DFT}}$ at $T = T_m^{\text{DFT}}$ and the relation $\frac{\partial G}{\partial T}= -S $ we obtain,
\begin{align}
\label{eq:eq01}
\Delta T_m = \frac{G_l^{\text{DFT}}(T_m^{\text{NNP}}) - G_s^{\text{DFT}}(T_m^{\text{NNP}})}{S_l^{\text{DFT}}(T_m^{\text{NNP}}) - S_s^{\text{DFT}}(T_m^{\text{NNP}})}.
\end{align}
By expressing $G^{\text{DFT}}$ and $S^{\text{DFT}}$ in terms of $G^{\text{NNP}}$ and $S^{\text{NNP}}$, respectively, and inserting in Eq.~\eqref{eq:eq01} we arrive at the final equation for $\Delta T_m$,
\begin{align}
\label{eq:eq02}
\Delta T_m = \frac{\Delta G_l - \Delta G_s}{S_l^{\text{NNP}} - S_s^{\text{NNP}} + \Delta S_l - \Delta S_s},
\end{align}
where $\Delta G = G^{\text{DFT}} - G^{\text{NNP}}$ and $\Delta S = S^{\text{DFT}} - S^{\text{NNP}}$. All quantities of Eq.~\eqref{eq:eq02} are evaluated at $T = T_m^{\text{NNP}}$. The difference $S_l^{\text{NNP}} - S_s^{\text{NNP}}$ is the entropy of fusion and is known from the interface pinning simulations (see Table~\ref{tab:density_liquid_solid}). With
\begin{align}
\Delta G =& \bigl< E \bigr>_{\text{DFT}} - \bigl< E \bigr>_{\text{NNP}} - T\Bigl( \bigl< S \bigr>_{\text{DFT}} - \bigl< S \bigr>_{\text{NNP}}\Bigr) \\\nonumber
	  & + p\Bigl( \bigl< V \bigr>_{\text{DFT}} - \bigl< V \bigr>_{\text{NNP}}\Bigr) = \Delta\overline{E} - T\Delta\overline{S} + p \Delta\overline{V}
\end{align}
we can find an expression for $\Delta S$,
\begin{align}
\label{eq:eq03}
\Delta S = \frac{1}{T} \Bigl( \Delta\overline{E} + p \Delta\overline{V} - \Delta G \Bigr).
\end{align}
Here $\bigl< \dots \bigr>_{\text{NNP}}$ and $\bigl< \dots \bigr>_{\text{DFT}}$ refer to averages corresponding to the NNP and the DFT potential-energy surface, respectively. Using thermodynamic perturbation theory, the averages $\Delta\overline{E}$, $\Delta\overline{V}$, and $\Delta G$ can be expressed as,
\begin{align}
\label{eq:eq04}
 \Delta\overline{E} =& \frac{\bigl<e^{-\beta\Delta E}E^{\text{DFT}}\bigr>_{\text{NNP}}}{\bigl<e^{-\beta\Delta E} \bigl>_{\text{NNP}}} - \bigl< E \bigr>_{\text{NNP}},\\\nonumber
 \Delta\overline{V} =& \frac{\bigl<e^{-\beta\Delta E}V^{\text{DFT}}\bigr>_{\text{NNP}}}{\bigl<e^{-\beta\Delta E} \bigr>_{\text{NNP}}} - \bigl< V \bigr>_{\text{NNP}},\\\nonumber
 \Delta G =& - k_{\text{B}}T\ln\bigl<e^{-\beta\Delta E} \bigl>_{\text{NNP}},
\end{align}
where $\Delta E = E^{\text{DFT}} - E^{\text{NNP}}$ and $\beta = 1 / k_{\text{B}}T$. The quantities needed to compute $\Delta T_m$ are obtained in the following way:
\begin{itemize}
 \item NNP simulations for the solid and the liquid phase are performed in the $NpT$ ensemble at $T = T_m^{\text{NNP}}$ and $p = 1$\,bar.\\[0.5pt]
 \item Independent configurations are extracted from the trajectories and their energies are recomputed with the corresponding DFT method in order to obtain $E^{\text{DFT}}$.\\[0.5pt]
 \item The averages $\bigl<e^{-\beta\Delta E}\bigr>_{\text{NNP}}$, $\bigl<e^{-\beta\Delta E}E^{\text{DFT}}\bigr>_{\text{NNP}}$, $\bigl<e^{-\beta\Delta E}V^{\text{DFT}}\bigr>_{\text{NNP}}$, $\bigl< E\bigr>_{\text{NNP}}$, and $\bigl< V \bigr>_{\text{NNP}}$ are computed both for the solid and the liquid phase.
 \item $\Delta G_l$, $\Delta G_s$, $\Delta S_l$, and $\Delta S_s$ are determined from Eqs.~\eqref{eq:eq04} and \eqref{eq:eq03}.
 \item $\Delta T_m$ is computed using Eq.~\eqref{eq:eq02}.
\end{itemize}

$NpT$ simulation for all NNPs were performed using 128 molecules for both phases and total simulation times of 14\,ns per NNP. After discarding 4\,ns for the purpose of equilibration, configurations were extracted every 20\,ps and their energies were recomputed with the corresponding reference DFT method. Fig.~\ref{fig:melting_point_correction}b-d shows the convergence of $\Delta G$, $\Delta S$, and $\Delta T_m$ with the number of configurations used to obtain the averages given in Eq.~\eqref{eq:eq04}. The final values for $T_m^{\text{DFT}}$, $T_m^{\text{NNP}}$, and $\Delta T_m$ are reported in Table~\ref{tab:melting_point_correction}. For all NNPs the correction term is positive, which originates from a positive shift of the NNP energies of the solid phase with respect to the DFT values (see Fig.~\ref{fig:melting_point_correction}e-h).

\begin{table}[htbp]
\centering
\caption{\label{tab:melting_point_correction} \textbf{Corrected melting temperatures.} Corrected melting temperatures $T_m^{\text{DFT}}$, melting temperatures obtained from interface pinning simulations $T_m^{\text{NNP}}$, and correction term $\Delta T_m$ obtained from thermodynamic perturbation theory calculations.
}
\begin{tabular}{llll}
 & \multicolumn{1}{c}{$T_m^{\text{DFT}}$(K)} & \multicolumn{1}{c}{$T_m^{\text{NNP}}$(K)} & \multicolumn{1}{c}{$\Delta T_m$(K)}	\\
\hline\noalign{\smallskip}
NNP(BLYP) 		& 323 $\pm$ 3 	& 321 $\pm$ 3	& 2.4	\\
NNP(BLYP-vdW) 	& 283 $\pm$ 2 	& 272 $\pm$ 2	& 11.4		\\
NNP(RPBE) 		& 267 $\pm$ 2 	& 265 $\pm$ 2	& 2.0	\\
NNP(RPBE-vdW) 	& 274 $\pm$ 3 	& 269 $\pm$ 3	& 5.2		\\
 \hline
\end{tabular}
\end{table}

\begin{table}[htbp]
\caption{\label{tab:density_liquid_solid} \textbf{Density of the liquid and the solid phase and entropy of fusion.} Density values of the liquid ($\rho_{\text{l}}$) and the solid ice I\textit{h} phase ($\rho_{\text{s}}$) together with the density difference ($\Delta\rho = \rho_{\text{l}} - \rho_{\text{s}}$) (in g/cm$^3$) and entropy of fusion $\Delta$S (in 10$^{-4}$eV/K) obtained from interface pinning simulations using different NNPs. Experimental values were taken from \cite{Haynes2015}.
}
\begin{tabular}{lllll}
\noalign{\smallskip}
Model & \multicolumn{1}{c}{$\rho_{\text{l}}$} & \multicolumn{1}{c}{$\rho_{\text{s}}$} & \multicolumn{1}{c}{$\Delta\rho$} & \multicolumn{1}{c}{$\Delta$S} \\
\hline \noalign{\smallskip}
NNP(BLYP)	& 0.752		& 0.840		& -0.088	& 2.48 	\\
NNP(BLYP-vdW)	& 1.053 	& 0.915		& 0.138		& 2.39 	\\
NNP(RPBE)	& 0.678		& 0.786		& -0.108	& 3.34 	\\
NNP(RPBE-vdW)	& 0.904		& 0.864		& 0.040		& 2.69 	\\
Exp.		& 0.99984 	& 0.91670	& 0.08314 	& 2.28	\\
\hline
\end{tabular}
\end{table}

\section{Neighbor Distribution}
The structure of water was analyzed by decomposing the oxygen-oxygen radial distribution function into contributions from neighboring molecules (similar to the analysis in Ref.~\cite{Saitta2003}) using the analysis tool TRAVIS~\cite{Brehm2011}. 
In order to visualize the location of first-shell and second-shell molecules, the centroid of the corresponding distribution functions 
$\text{P}_{1^{\mathrm{st}}-4^{\mathrm{th}}}\left(r_{\text{OO}}\right)$
and 
$\text{P}_{5^{\mathrm{th}}-8^{\mathrm{th}}}(r_{\text{OO}})$ 
was computed, and the position of the centroid,
\begin{align}
 C_x = \frac{\int \text{P}(r_{\text{OO}}) r_{\text{OO}}\,\,dr_{\text{OO}}}{\int \text{P}(r_{\text{OO}})\,\,dr_{\text{OO}}},
\end{align}
was plotted against temperature. Error bars for $C_x$ were estimated by averaging over values obtained from non-overlapping time intervals of a length of 0.5\,ns.

\section{Hydrogen Bond Fluctuation}

\begin{figure*}[htbp]
\centering
  \includegraphics{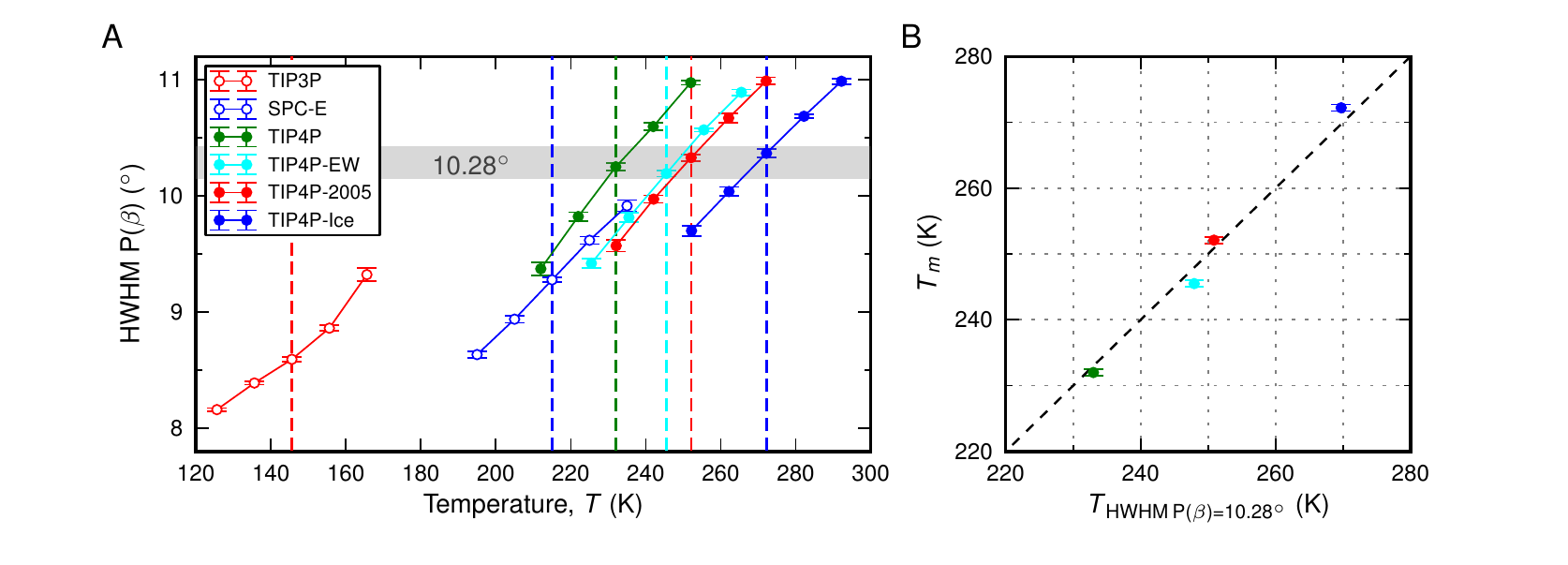}\\  
\caption{\label{fig:hbond_melt}\textbf{Hydrogen bond strength and melting temperature for common empirical water models.} (A) Fluctuation in the hydrogen bond angle $\beta$ close to the melting temperature of ice I\textit{h}~\cite{Vega2005a}, obtained from simulations using the six empirical water models~\cite{Jorgensen1983,Berendsen1987,Horn2004,Abascal2005,Abascal2005a} given in the legend box. Dashed vertical lines denote the melting temperature of the respective water model. (B) Correlation between melting temperature \textit{T}$_\text{\textit{m}}$ and temperature of a hydrogen bond fluctuation of 10.28$^\circ$ computed for water models based on four-site interaction potentials (TIP4P-\textit{x}).}
\end{figure*}

HB strengths were measured in terms of fluctuations in the distribution P($\beta$) of the HB angle $\beta=\angle\text{H}_\text{D}\!\!-\!\text{O}_\text{D}\!\cdots\!\text{O}_\text{A}$. Cone corrected~\cite{Kroon1974} probability density functions P($\beta$) were obtained by computing $\beta$ between a reference molecule and its four nearest neighbors. Fluctuations were obtained from the half width at half maximum (HWHM) of a Gaussian fit to the probability density function. Error bars for HWHM P$(\beta)$ were estimated by averaging over values obtained from non-overlapping time intervals of a length of 0.5\,ns. Angular probability density functions P($\beta$) and combined angular/radial probability density functions were computed using TRAVIS~\cite{Brehm2011}. 
Molecular dynamics simulations using six empirical water models (TIP3P~\cite{Jorgensen1983}, SPC/E~\cite{Berendsen1987}, TIP4P~\cite{Jorgensen1983}, TIP4P-Ew~\cite{Horn2004}, TIP4P/2005~\cite{Abascal2005}, and TIP4P/Ice~\cite{Abascal2005a}) at regions close to their melting temperature~\cite{Vega2005a} were carried out to confirm the correlation between hydrogen bond strength and melting temperature. These simulations were performed using 2880 water molecules with a time step of $\Delta t =$ \,2\,fs. At each temperature, a trajectory of 5\,ns length was generated to extract hydrogen bond information. The fluctuations of the hydrogen bond angle for the empirical water models are depicted in Fig.~\ref{fig:hbond_melt}.

\section{van der Waals Interactions}

\begin{figure*}[htbp]
\centering
  \includegraphics{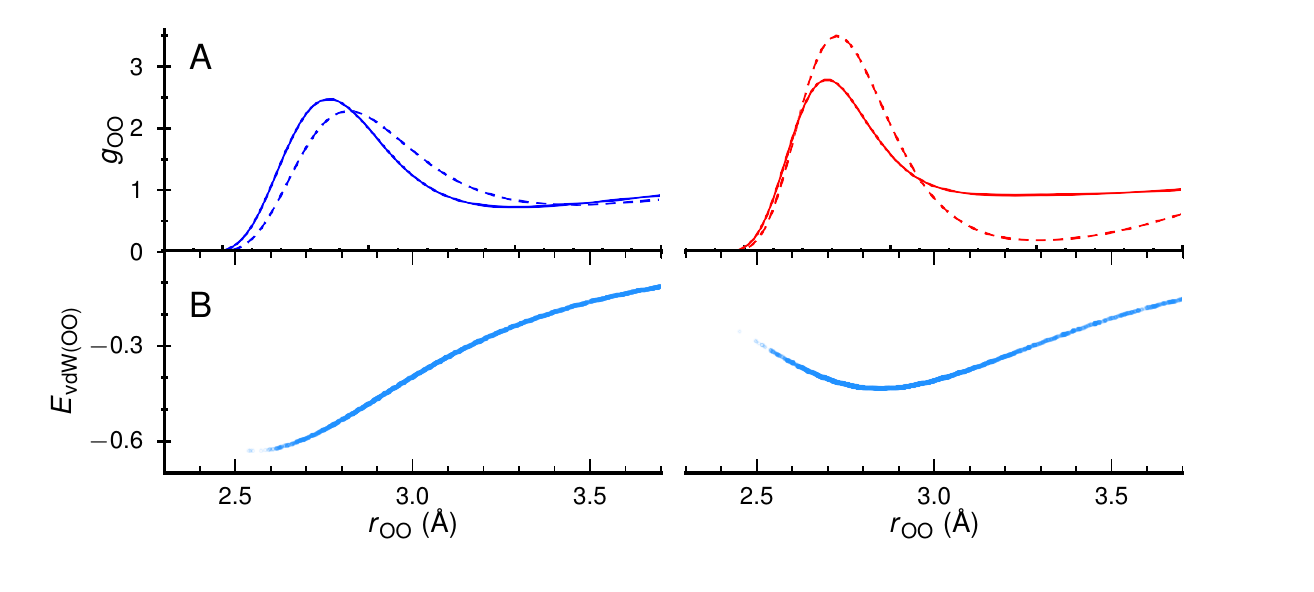}\\  
\caption{\label{fig:vdW_OO}\textbf{Effect of van der Waals interactions on oxygen-oxygen distributions.}
(A) Oxygen-oxygen radial distribution functions \textit{g}$_\text{OO}$ from NNP simulations at 300\,K based on the RPBE (left) and BLYP (right) density-functionals with (solid lines) and without (dashed lines) vdW corrections. 
(B) Van der Waals pair interaction energy \textit{E}$_{\text{vdW(OO)}}$ (in $k_{\rm B}T$) between pairs of oxygen atoms as a functions of the pair distance \textit{r}$_{\text{OO}}$ obtained from the NNP simulations. 
}
\end{figure*}

We employed the DFT-D3 method~\cite{Grimme2010} in order to account for vdW interactions. In this approach the two-body van der Waals interaction $E_{\text{vdW}}^{(2)}$ for atom pairs $AB$ at distance $r_{AB}$ is computed from 6$^{\mathrm{th}}$- and 8$^{\mathrm{th}}$-order dispersion coefficients $C^{AB}_{6/8}$ that depend on their chemical environment (by being a function of fractional coordination numbers $CN$, cf. Ref.~\cite{Grimme2010}),
\begin{align}
 E_{\text{vdW}}^{(2)} =& -\sum_{A<B}^{N_{\text{pairs}}} \biggl( \frac{C_6^{AB}(CN)}{r^6_{AB}}f_{d,6}(r_{AB}) \\\nonumber
 &+s_8\frac{C_8^{AB}(CN)}{r^8_{AB}}f_{d,8}(r_{AB}) \biggr).
\end{align}
The range of the vdW correction is determined by damping functions $f_{d,n}$, which screen the vdW contribution to zero at short distances (zero-damping) in order to avoid near singularities,
\begin{align}
 f_{d,n}(r_{AB}) = \Biggl( 1 + 6\biggl( \frac{r_{AB}}{s_{r,n}R_0^{AB}} \biggr)^{-\alpha_n} \Biggr)^{-1}.
\end{align}
The parameters $s_8$ and $s_{r,6}$ are the only two density-functional dependent parameters of the D3 method (cf. Table~\ref{tab:vdW_coeffs}). Van der Waals pair interactions, \textit{E}$_{\text{vdW(OH/OO)}}$, for oxygen-hydrogen and oxygen-oxygen pairs (shown in Fig.~\ref{fig:fig05} of the main text and in Fig.~\ref{fig:vdW_OO}, respectively), and average values of $C^{AB}_{6/8}$ coefficients reported in Table~\ref{tab:vdW_coeffs} were computed by employing a modified version of the dftd3 program~\cite{GrimmeD3}. 

\begin{table}[htbp]
\centering
\caption{\label{tab:vdW_coeffs}\textbf{Van der Waals coefficients and density-functional dependent parameters.} Environment-dependent van der Waals coefficients for oxygen-hydrogen, oxygen-oxygen, and hydrogen-hydrogen pairs, $C^{AB}_{6/8}$ (in a.u.), averaged over trajectories from \textit{NpT} simulations at 300\,K based on BLYP and RPBE, respectively (standard deviation is given in parentheses) and values for the two density-functional dependent parameters of the D3 method ($s_{r,6}$ and $s_8$).}
\begin{tabular}{lrr}
 & \multicolumn{1}{c}{BLYP} & \multicolumn{1}{c}{RPBE} 	\\
\hline\noalign{\smallskip}
$C_6^{\text{OH}}$ 	& 5.436    (0.004)	& 5.437    (0.003)	\\
$C_8^{\text{OH}}$ 	& 84.897   (0.062)	& 84.922   (0.042)	\\
$C_6^{\text{OO}}$ 	& 10.410   (0.003)	& 10.413   (0.002)	\\
$C_8^{\text{OO}}$ 	& 210.087  (0.067)	& 210.134  (0.046)	\\
$C_6^{\text{HH}}$ 	& 3.092    (0.003)	& 3.093    (0.002)	\\
$C_8^{\text{HH}}$ 	& 37.382   (0.038)	& 37.395   (0.026)	\\
$s_{r,6}$		& 1.094			& 0.872			\\
$s_8$			& 0.722			& 0.514			\\
 \hline
\end{tabular}
\end{table}

As shown in Table~\ref{tab:vdW_coeffs}, for both density-functionals the $C^{AB}_{6/8}$ coefficients are essentially identical, responsible for the different range of the vdW pair interaction are solely the $s_{r,6}$ and $s_8$ parameters.

\end{document}